\documentclass[11pt]{scrartcl}

\newcommand{\blitza}{\text{\usefont{U}{ulsy}{m}{n}\symbol{'011}}}

\usepackage{graphicx,framed,subfigure}
\usepackage{wrapfig,fullpage}
\usepackage{enumitem}

\usepackage{amsmath,amssymb,latexsym,amsopn,amsfonts}
\usepackage[boxed,ruled,vlined,linesnumbered,noresetcount]{algorithm2e}
\makeatletter
\newcommand{\nosemic}{\renewcommand{\@endalgocfline}{\relax}}
\newcommand{\dosemic}{\renewcommand{\@endalgocfline}{\algocf@endline}}
\newcommand{\pushline}{\Indp}
\newcommand{\popline}{\Indm\dosemic}
\let\oldnl\nl
\newcommand{\nonl}{\renewcommand{\nl}{\let\nl\oldnl}}
\makeatother

\newcommand\bull{{\operatorname{-\xspace}}}

\usepackage{cite}
\usepackage{amsmath,amssymb,amsfonts}
\usepackage{graphicx}
\usepackage{textcomp}
\usepackage{xcolor}
\def\BibTeX{{\rm B\kern-.05em{\sc i\kern-.025em b}\kern-.08em
    T\kern-.1667em\lower.7ex\hbox{E}\kern-.125emX}}

\newcommand{\ems}[1]{#1}
\definecolor{oskar_green}{rgb}{0.0, 0.5, 0.0}

\newcommand{\algSize}{normalsize} 

\newcommand{\B}{\vspace*{-\smallskipamount}}

\newcommand{\F}{\vspace*{\smallskipamount}}
\newcommand{\FF}{\vspace*{\medskipamount}}

\newcommand{\bigO}{\mathcal{O}\xspace}
\newcommand{\remove}[1]{}
\newcommand{\reduce}[1]{#1}

\newcommand{\txD}{\mathit{txDes}\xspace} 
\newcommand{\exist}{\mathsf{hasTerminated}\xspace}
\newcommand{\nonActive}{\mathsf{allHaveTerminated}\xspace}
\newcommand{\done}{\mathsf{result}\xspace}

\newcommand{\test}{\mathsf{test}\xspace} 

\newcommand{\fifoReady}{\mathsf{readyMax}\xspace} 
\newcommand{\retrieve}{\mathsf{readyMin}\xspace}
\newcommand{\bulkRead}{\mathsf{bulkRead}\xspace}
\newcommand{\getSeq}{\mathsf{getSeq}\xspace}
\newcommand{\7}{{3}} 
\newcommand{\6}{{2}} 
\newcommand{\5}{{1}} 

\newcommand{\minReady}{\mathit{minReady}\xspace}
\newcommand{\maxReady}{\mathit{maxReady}\xspace}

\newcommand{\Sset}{\mathsf{S}\xspace}

\newcommand{\exceed}{\Delta}

\newcommand{\etal}{\emph{et al.}\xspace}
\newcommand{\eg}{\emph{e.g.,}\xspace}
\newcommand{\Eg}{\emph{E.g.,}\xspace}
\newcommand{\ie}{\emph{i.e.,}\xspace}
\newcommand{\Ie}{\emph{I.e.,}\xspace}

\newcommand{\xS}{\mathit{obsS}\xspace}

\newcommand{\minObstSeq}{\mathsf{minObsS}\xspace}

\newtheorem{theorem}{Theorem}[section]
\newtheorem{lemma}[theorem]{Lemma}
\newtheorem{definition}{Definition}[section]

\newtheorem{assumption}[theorem]{Assumption}
\newtheorem{claim}[theorem]{Claim}
\newcommand{\true}{\mathsf{True}\xspace}

\newcommand{\false}{\mathsf{False}\xspace}

\newenvironment{claimProof}{\par\noindent\textbf{Proof of Claim  \thmcnt\space}}{\hfill $\Box_{Claim ~ \clmcnt}$}
\newenvironment{lemmaProof}{\par\noindent\textbf{Proof of Lemma  \thmcnt\space}}{\hfill $\Box_{Lemma ~ \lemcnt}$}
\newenvironment{theoremProof}{\par\noindent\textbf{Proof of Theorem  \thmcnt\space}}{\hfill $\Box_{Theorem ~ \thmcnt}$}
\newcommand{\clmcnt}{0}
\newcommand{\lemcnt}{0}
\newcommand{\thmcnt}{0}

\newcommand{\sP}{\mathcal{P}\xspace}

\newcommand{\N}{\mathbb{N}\xspace}
\newcommand{\capacity}{\mathsf{channelCapacity}\xspace}

\newcommand{\Correct}{\mathit{Correct}\xspace}

	\newcommand{\Section}[1]{\section{#1}}
	\newcommand{\Subsection}[1]{\subsection{#1}}
	\newcommand{\Subsubsection}[1]{\subsubsection{#1}}

\remove{ 
\newcommand{\Section}[1]{\section{#1}\B}
\newcommand{\Subsection}[1]{\noindent \textbf{#1.}~~}
\newcommand{\Subsubsection}[1]{\noindent \emph{#1.}~}

} 

\begin{document}

\title{Self-stabilizing Multivalued Consensus\\ in Asynchronous Crash-prone Systems\\\FF\Large{(preliminary version)}
}


\author{Oskar Lundstr\"om \and Michel Raynal \and Elad M.\ Schiller}

\maketitle

\begin{abstract}
The problem of multivalued consensus is fundamental in the area of fault-tolerant distributed computing since it abstracts a very broad set of agreement problems in which processes have to uniformly decide on a specific value $v \in V$, where $|V| \geq 2$. Existing solutions (that tolerate process failures) reduce the multivalued consensus problem to the one of binary consensus, \eg Most{\'{e}}faoui-Raynal-Tronel and Zhang-Chen. 

Our study aims at the design of an even more reliable  solution. We do so through the lenses of \emph{self-stabilization}---a very strong notion of fault-tolerance. In addition to node and communication failures, self-stabilizing algorithms can recover after the occurrence of \emph{arbitrary transient-faults}; these faults represent any violation of the assumptions according to which the system was designed to operate (as long as the algorithm code stays intact).  

This work proposes the first (to the best of our knowledge) self-stabilizing algorithm for multivalued consensus for asynchronous message-passing systems prone to process failures and arbitrary transient-faults. Our solution is also the first (to the best of our knowledge) to support wait-freedom. Moreover, using piggybacking techniques, our solution can invoke $n$ binary consensus objects concurrently. Thus, the proposed self-stabilizing solution can terminate using fewer binary consensus objects than earlier non-self-stabilizing solutions by Most{\'{e}}faoui, Raynal, and Tronel, which uses an unbounded number of binary consensus objects, or Zhang and Chen, which is not wait-free.


\end{abstract}


\Section{Introduction}
\label{sec:intro}
We propose, to the best of our knowledge, the first self-stabilizing, non-blocking, and memory-bounded implementation of \emph{multivalued consensus} objects for asynchronous message-passing systems whose nodes may crash.

\Subsection{Background and motivation} 
Fault-tolerant distributed applications span over many domains in the area of banking, transports, tourism, production, commerce, to name a few. The implementations of these applications use message-passing systems and require fault-tolerance. The task of designing and verifying these systems is known to be very hard, because the joint presence of failures and asynchrony creates uncertainties about the application state (from the process's point of view). \Eg Fischer, Lynch, and Paterson~\cite{DBLP:journals/jacm/FischerLP85} demonstrated that, in any asynchronous message-passing system, it takes no more than one process crash to prevent the system from achieving consensus deterministically.

Our focal application is the emulation of finite-state machines. For the sake of consistency maintenance, all emulating processes need to apply identical sequences of state transitions. This can be done by dividing the problem into two: (i) propagation of user input to all emulating processes, and (ii) letting each emulating process execute identical sequences of state transitions. Uniform reliable broadcast~\cite{DBLP:books/sp/Raynal18,hadzilacos1994modular} can solve Problem (i). This work focuses on Problem (ii) since it is the core problem. \Ie all processes need to agree on a common value according to which all emulating processes execute their state transitions.
The consensus problem generalizes problem (ii) and requires each process to propose a value, and all non-crashed processes to reach a common decision that one of them had proposed. There is a rich literature on fault-tolerant consensus. This work advances the state of the art by offering a greater set of failures that can be tolerated.

\begin{figure}
	\begin{center}
		\includegraphics[scale=0.45, clip]{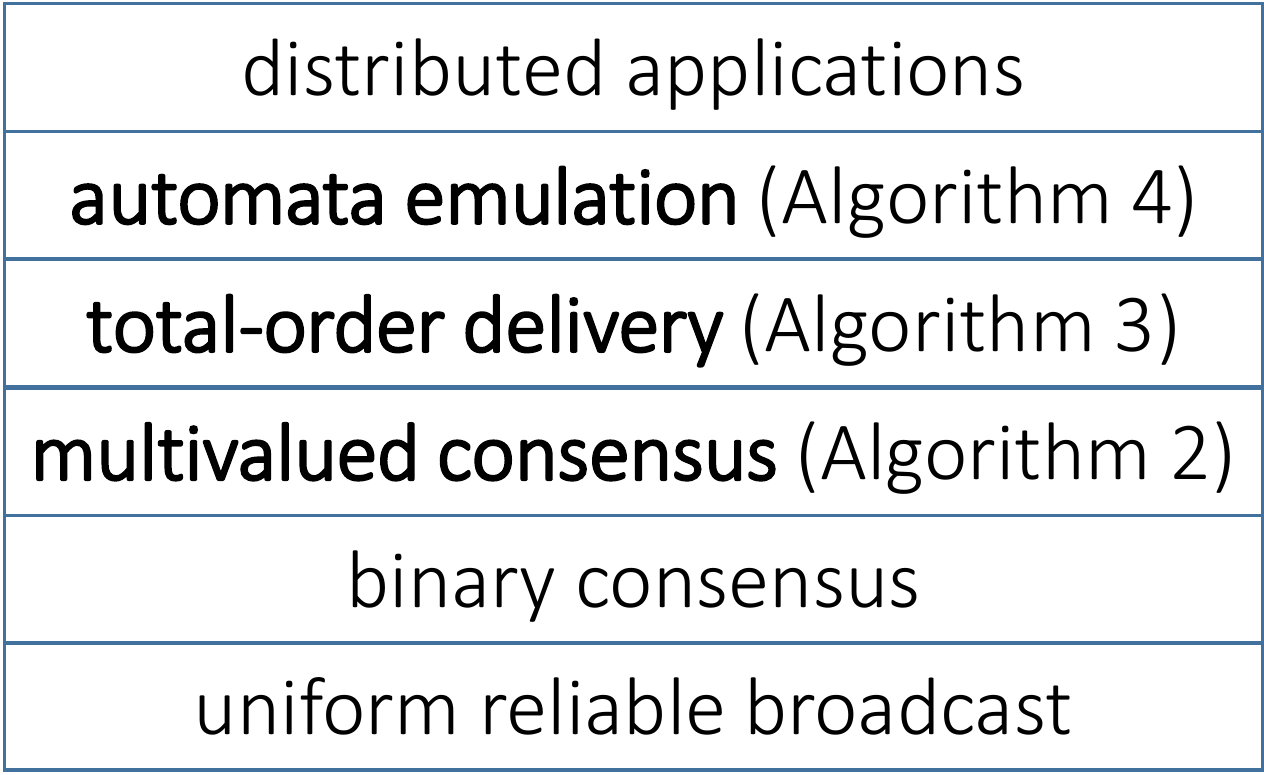}
	\end{center}
	\caption{\label{fig:suit}{The studied problems\reduce{ of binary consensus} (in bold font) and their context}}
\end{figure}

\Subsection{Problem definition and scope} 
The definition of the consensus problem appears in Definition~\ref{def:consensus}. This work studies the multivalued version of the problem in which there are at least two values that can be proposed. Note that there is another version of the problem in which this set includes exactly two values, and referred to as binary consensus. Existing solutions for the multivalued consensus (as well as the proposed one) often use binary consensus algorithms. We present the relation among the problems mentioned above in Figure~\ref{fig:suit}.    

\remove{The consensus problem}

\begin{definition}[Consensus]
	\label{def:consensus}
	Every process $p_i$ has to propose a value $v_i \in V$ via an invocation of the $\mathsf{propose}_i(v_i)$ operation, where $V$ is a finite set of values. Let $\mathit{Alg}$ be an algorithm that solves consensus. $\mathit{Alg}$ has to satisfy \emph{safety} (\ie validity, integrity, and agreement) and \emph{liveness} (\ie termination).
	\begin{itemize}[leftmargin=0.15in]
		\item \textbf{Validity.} Suppose that $v$ is decided. Then some process had invoked $\mathsf{propose}(v)$.
		\item \textbf{Termination.} All non-faulty processes decide.
		\item \textbf{Agreement.} No two processes decide different values.
		\item \textbf{Integrity.} No process decides more than once.
	\end{itemize}
\end{definition}


\Subsection{Fault Model} 
We consider an asynchronous message-passing system that has no guarantees on communication delays (except that they are finite) and the algorithm cannot explicitly access the local clock. Our fault model includes $(i)$ crashes of less than half of the processes, and $(ii)$ communication failures, such as packet omission, duplication, and reordering. 

In addition to the failures captured in our model, we also aim to recover from \emph{arbitrary transient-faults}, \ie any temporary violation of assumptions according to which the system and network were designed to operate, \eg the corruption of control variables, such as the program counter, packet payload, and indices, \eg sequence numbers, which are responsible for the correct operation of the studied system, as well as operational assumptions, such as that at least a majority of nodes never fail. Since the occurrence of these failures can be arbitrarily combined, it follows that these transient-faults can alter the system state in unpredictable ways. In particular, when modeling the system, we assume that these violations bring the system to an arbitrary state from which a \emph{self-stabilizing algorithm} should recover the system after the occurrence of the last transient-fault. The system is guaranteed to satisfy the task requirements, \eg Definition~\ref{def:consensus}, after this recovery period. Our design criteria also support wait-freedom, which requires all operations to terminate within a bounded number of algorithm steps. Wait-freedom is important since it assures starvation-freedom even in the presence of failures since all operations terminate (as long as the process that invoked them does not crash).

\Subsection{Related Work} 
The celebrated Paxos algorithm~\cite{DBLP:journals/tocs/Lamport98} circumvents the impossibility by Fischer, Lynch, and Paterson~\cite{DBLP:journals/jacm/FischerLP85}, from now on FLP, by assuming that failed computers can be detected by unreliable failure detectors~\cite{DBLP:journals/jacm/ChandraT96}. Paxos has inspired many veins of research, \eg~\cite{DBLP:journals/csur/RenesseA15} and references therein. We, however, follow the family of abstractions by Raynal~\cite{DBLP:books/sp/Raynal18} due to its clear presentation that is easy to grasp. Also, the studied algorithm does not consider failure detectors. Instead, it assumes the availability of binary consensus objects, which uses the weakest failure detector, see Raynal~\cite{DBLP:books/sp/Raynal18}.  

\Subsubsection{Non-self-stabilizing solutions} 
Most{\'{e}}faoui, Raynal, and Tronel~\cite{DBLP:journals/ipl/MostefaouiRT00}, from now on MRT, reduce multivalued consensus to binary consensus via a crash-tolerant block-free algorithm.
%
%
MRT uses an unbounded number of invocations of binary consensus objects and at most one uniform reliable broadcast (URB) per process. 
Zhang and Chen~\cite{DBLP:journals/ipl/ZhangC09a} proposed an algorithm for multivalued consensus that uses only $x$ instances, where $x$ is the number of bits it takes to represent any value in $V$; the domain of proposable values. 

Our self-stabilizing solution is wait-free since termination is achieved within at most $n$ invocations of binary consensus objects and at most one uniform reliable broadcast~\cite{selfStabURB} (URB) operation per process, where $n$ is the number of processes in the system. 
However, each such URB invocation needs to be repeated until the consensus object is deactivated by the invoking algorithm. This is due to a well-known impossibility~\cite[Chapter 2.3]{DBLP:books/mit/Dolev2000}, which says that self-stabilizing systems cannot terminate and stop sending messages. Note that it is easy to trade the broadcast repetition rate with the speed of recovery from transient-faults.

Afek \etal~\cite{DBLP:journals/dc/AfekGRRT10} showed that binary and multivalued versions of the $k$-simultaneous consensus task are wait-free equivalent. Here, the $k$-simultaneous consensus is required to let each process to participate at the same time in $k$ independent consensus instances until it decides in any one of them.

\reduce{Our study focuses on deterministic solutions and does not consider probabilistic approaches, such as~\cite{DBLP:conf/isorc/EzhilchelvanMR01,DBLP:conf/podc/LiangV11,DBLP:journals/cj/BabaeeD14}. It is worth mentioning that Byzantine fault-tolerant multivalued consensus algorithms~\cite{DBLP:conf/nca/CrainGLR18,DBLP:journals/acta/MostefaouiR17,DBLP:journals/cj/CorreiaNV06,DBLP:conf/sirocco/Tseng16} have applications to Blockchain~\cite{DBLP:conf/sirocco/MostefaouiR15}. Our fault model does not include Byzantine failures, instead, we consider arbitrary transient-faults.}

\Subsubsection{Self-stabilizing solutions} 
We follow the design criteria of self-stabilization, which Dijkstra~\cite{DBLP:journals/cacm/Dijkstra74} proposed. A detailed pretension of self-stabilization was provided by Dolev~\cite{DBLP:books/mit/Dolev2000} and Altisen \etal~\cite{DBLP:series/synthesis/2019Altisen}. 
Consensus was sparsely studied in the context of self-stabilization. Blanchard \etal~\cite{DBLP:conf/netys/BlanchardDBD14} presented the first solution in the context of self-stabilization. They presented a practically-self-stabilizing version of Paxos~\cite{DBLP:journals/tocs/Lamport98}, which was the first (non-self-stabilizing) solution to the area of fault-tolerant message-passing systems. The studied solution is part of a more advanced and efficient protocol suite (Figure~\ref{fig:suit}). 
We note that practically-self-stabilizing systems, as defined by Alon \etal~\cite{DBLP:journals/jcss/AlonADDPT15}\reduce{ and clarified by Salem and Schiller~\cite{DBLP:conf/netys/SalemS18}}, do not satisfy Dijkstra's requirements, \ie practically-self-stabilizing systems do not guarantee recovery within a finite time after the occurrence of transient-faults. 
We base our self-stabilizing multivalued consensus on the self-stabilizing binary consensus by Lundstr{\"{o}}m, Raynal, and Schiller~\cite{DBLP:conf/icdcn/LundstromRS21}, which is the first self-stabilizing solution to the binary consensus problem that recovers within a bounded time.


We propose, to the best of our knowledge, the first self-stabilizing solution for the multivalued version of the problem. As an application, we offer, to the best of our knowledge, the first self-stabilizing algorithm for (uniform reliable broadcast with) total order delivery. It is based on the self-stabilizing uniform reliable broadcast with FIFO delivery by Lundstr{\"{o}}m, Raynal, and Schiller~\cite{selfStabURB}. Our solution can facilitate the self-stabilizing emulation of state-machine replication. Dolev \etal~\cite{DBLP:journals/jcss/DolevGMS18} proposed the first practically-self-stabilizing emulation of state-machine replication, which has a similar task to one in Figure~\ref{fig:suit}. However, Dolev \etal's solution does not guarantee recovery within a finite time since it does not follow Dijkstra's criterion. Moreover, it is based on virtual synchrony by Birman and Joseph~\cite{DBLP:journals/tocs/BirmanJ87}, where the one in Figure~\ref{fig:suit} considers censuses. 




Georgiou, Lundstr{\"{o}}m, and Schiller studied the trade-off between non-blocking and wait-free solutions for self-stabilizing atomic snapshot objects~\cite{DBLP:conf/netys/GeorgiouLS19}. We study a similar trade-off for a different problem.

More generally, in the context of self-stabilization there are algorithms for group communications~\cite{DBLP:journals/tpds/DolevS03,DBLP:journals/acta/DolevS04,DBLP:journals/tmc/DolevSW06}, consensus in shared-memory systems~\cite{DBLP:journals/jcss/DolevKS10}, wireless communications~\cite{DBLP:journals/ijdsn/LeoneS10,DBLP:journals/tcs/HoepmanLST11,DBLP:conf/vtc/MustafaPSTT12,DBLP:journals/ijdsn/LeoneS13,DBLP:conf/sss/PetigST13,DBLP:conf/medhocnet/PetigST14,DBLP:journals/sensors/PonceSF18}, software defined networks~\cite{DBLP:conf/icdcs/CaniniSSSS18,DBLP:conf/icdcs/CaniniSSSS17}, virtual infrastructure for mobile nodes~\cite{DBLP:conf/dialm/DolevGSSW05,DBLP:conf/wdag/DolevGLSSW04,DBLP:conf/eurongi/WegenerSHFF06}, to name a few.

%

\Subsection{Our contribution} 
We present a fundamental module for dependable distributed systems: a self-stabilizing wait-free algorithm for multivalued consensus for asynchronous message-passing systems that are prone to crash failures. To the best of our knowledge, we are the first to provide a wait-free solution for multivalued consensus that tolerates a broad fault model \ie crashes, communication failures, \eg packet omission, duplication, and reordering as well as arbitrary transient-faults using a bounded amount of resources. The latter models any violation of the assumptions according to which the system was designed to operate (as long as the algorithm code stays intact). 

Our wait-free solution achieves (multivalued) consensus within $n$ invocations of binary consensus instances that can run either sequentially or concurrently, where $n$ is the number of processes. Besides, our concurrent version can piggyback the binary consensus messages and terminate within the time that it takes to complete one uniform reliable broadcast (URB) and one binary consensus. This is also the time it takes the system to recover after the occurrence of the last transient-fault. 

As an application, this technical report offers a total order extension to the self-stabilizing FIFO URB service by Lundstr{\"{o}}m, Raynal, and Schiller~\cite{selfStabURB}. That self-stabilizing solution uses three multivalued consensus objects and stabilizes within a constant time. The technical report also explains how to enhance this solution to a self-stabilizing emulator of a replicated state machine.     


\Subsection{Organization} We state our system settings in Section~\ref{sec:sys}. The task specifications and solution organization appear in Section~\ref{sec:omega}. Section~\ref{sec:back} includes a brief overview of the studied algorithm by Most{\'{e}}faoui, Raynal, and Tronel~\cite{DBLP:journals/ipl/MostefaouiRT00} that has led to the proposed solution. Our self-stabilizing algorithm for consensus multivalued object is proposed in Section~\ref{sec:seflStab}. The correctness proof appears in Section~\ref{sec:corr}. We present an application to the proposed algorithm in Section~\ref{sec:bounded}, which is a self-stabilizing total order uniform reliable broadcast. We conclude in Section~\ref{sec:disc} and explain how to extend the proposed application to serve as an emulator for state-machine replication.

%

\smallskip

\Section{Task Specifications and Solution Organization}
\label{sec:omega}

The proposed solution is tailored for the protocol suite presented in Figure~\ref{fig:suit}. Thus, before we specify how all these tasks are organized into one solution, we list the external building blocks and define the studied tasks.

\Subsection{External Building-Blocks: Uniform Reliable Broadcast}
\label{sec:blocks}

\Subsubsection{Binary consensus objects}
\label{sec:binCon}
$T_{\text{binCon}}$ denotes the task of binary and multivalued consensus, which Definition~\ref{def:consensus} specifies.
%
%
We assume the availability of self-stabilizing binary consensus objects, such as the one by Lundstr{\"{o}}m, Raynal, and Schiller~\cite{DBLP:conf/icdcn/LundstromRS21}.
As in Definition~\ref{def:consensus}, the proposed and decided values have to be from the $V$ domain (of proposable values). For clarity sake, we distinguish the invocation of binary and multivalued consensus. That is, for a given binary consensus object $BC$, the operation $BC.\mathsf{binPropose}(v)$ invokes the binary consensus on $ v \in V=\{\true,\false\}$. (Traditionally, the result of binary consensus is either $0$ or $1$, but we rename them.)

\Subsubsection{First-in first-out uniform reliable broadcast (FIFO-URB)}
\label{sec:URB}
The task $T_{URB}$ of Uniform reliable broadcast (URB)~\cite{hadzilacos1994modular} considers an operation for URB broadcasting of message $m$ and an event of URB delivery of message $m$. The requirements include URB-validity, \ie there is no spontaneous creation or alteration of URB messages, URB-integrity, \ie there is no duplication of URB messages, as well as URB-termination, \ie if the broadcasting node is non-faulty, or if at least one receiver URB-delivers a message, then all non-failing nodes URB-deliver that message. Note that the URB-termination property considers both faulty and non-faulty receivers. This is the reason why this type of reliable broadcast is named \emph{uniform}. 


%
\label{sec:fifoURB}
The task of FIFO-URB, denoted by $T_{\text{TO-URB}}$, requires, in addition to the above URB requirements, \ie URB-validity, URB-integrity, and URB-termination, that all messages that come from the same sender are delivered in the same order in which their sender has sent them; but there are no constraints regarding messages that arrive from different senders.
%
%

The proposed solution assumes the availability of a self-stabilizing uniform reliable broadcast (URB)~\cite{selfStabURB}. We also assume that the operation for URB broadcasting message $m$ returns a transmission descriptor, $\txD$, which is the unique message identifier. Moreover, the predicate $\exist(\txD)$ holds whenever the sender knows that all non-failing nodes in the system have delivered $m$. The implementation of $\exist(\txD)$ can just test that all trusted receivers have acknowledged the arrival of the message with identifier $\txD$. The solution in~\cite{selfStabURB} can facilitate the implementation of $\exist()$ since the self-stabilizing algorithm in~\cite{selfStabURB} considers such messages as 'obsolete' messages and lets the garbage collector remove them.

\Subsection{Task specifications}
\label{sec:spec}
We specify the studied tasks.

\Subsubsection{Total order URB (TO-URB)}
\label{sec:TOURB}
The task of total order URB, denoted by $T_{\text{TO-URB}}$, requires the total order delivery requirement, in addition to URB-validity, URB-integrity, and URB-termination. The total order delivery requirement says that if a node calls $\mathsf{toDeliver}(m)$ and later $\mathsf{toDeliver}(m')$, then no node $\mathsf{toDeliver}(m')$ before $\mathsf{toDeliver}(m)$. 

\Subsubsection{Binary and multivalued consensus objects}
$T_{\text{mulCon}}$ denotes the task of multivalued consensus, which Definition~\ref{def:consensus} specifies.
%
%
As in Definition~\ref{def:consensus}, the proposed and decided values have to be from the $V$ domain (of proposable values), where $|V|>2$. The operation $\mathsf{propose}(v)$ invokes the multivalued consensus on $ v \in V$.

\Subsection{Solution organization}
\label{sec:org}
We consider multivalued consensus objects that use an array, $BC[]$, of $n$ binary consensus objects, such as the one by~\cite[Chapter 17]{DBLP:books/sp/Raynal18}, where $n=|\sP|$ is the number of nodes in the system. The proposed algorithm considers a single multivalued consensus object, denoted by $O$. 

The proposed application, which is a TO-URB solution, considers an array, $CS[]$, of $M$ multivalued consensus objects, where $M \in \mathbb{Z}^+$ is a predefined constant. Our solution for TO-URB uses $M=3$ (Section~\ref{sec:bounded}). Each object is uniquely identified using a single sequence number. The proposed algorithm assumes that the multivalued consensus object $O$ is stored at $CS[s \bmod M]$. Whenever an operation is invoked or a message is sent, the sequence number $s$ is attached as a procedure parameter, and respectively, a message field (although the code of the proposed algorithm does not show this).  
We note that in case the proposed application runs out of sequence numbers, a global restart mechanism can be invoked, such as the one in~\cite[Section~5]{DBLP:conf/netys/GeorgiouLS19}. The function $\test(s)$ is used to assert consistency of the sequence number $s$. The function returns $\false$ whenever inconsistency is detected. Our TO-URB solution exemplifies an implementation of $\test()$. We assume that all underlying algorithms invoke $\test(s)$ whenever the object $CS[s \bmod M]$ or $CS[s \bmod M].BC[k]:p_k \in \sP$ is accessed, an operation is invoked, or a message (that is associated with $s$) arrives (although the code of the proposed algorithm does not show this). (The term underlying algorithm refers to both the proposed algorithm for multivalued consensus as well as the one for binary consensus.) If object $CS[s \bmod M]$ is found to be inconsistent, it is simply deactivated by assigning $\bot$ to $CS[s \bmod M]$. Also, inconsistent operation invocations and arriving messages are simply ignored.

Definition~\ref{def:consensus} considers the $\mathsf{propose}(v)$ operation, but it does not specify how the decided value is retrieved. We clarify that it can be either via the returned value of the $\mathsf{propose}(v)$ (or $\mathsf{binPropose}(v)$) operation (as in algorithm~\cite{DBLP:journals/tc/GuerraouiR04}) or via the returned value of the $\done()$ operation (as in the proposed solution). But, if $p_i \in \sP$ is yet to have access to the decided value, $\done_i()$ returns $\bot$. Otherwise, the decided value is returned. Specifically, for the case of $\mathsf{propose}(v)$, the parameter $s$ should be used when calling $\done_i(s)$ and for the case of $\mathsf{binPropose}(v)$, also the parameter $k:p_k \in \sP$ should be used when calling  $\done_i(s,k)$. 

We clarify that, in the absence of transient-faults, $\done_i(s)$ and $\done_i(s,k)$ always return either $\bot$ or the decided value. Thus, we solve the problem specified by Definition~\ref{def:consensus}. The studied algorithm~\cite{DBLP:journals/tc/GuerraouiR04} was not designed to deal with  transient-faults. As we explain in Section~\ref{sec:seflStabRes}, transient-faults can cause the studied algorithm to violate Definition~\ref{def:consensus}'s requirements without providing any indication to the invoking algorithm. After the occurrence of a transient-fault, the proposed solution allows $\done_i(s)$ to provide such indication to the invoking algorithm via the return of the \emph{transient error} symbol $\blitza$. Section~\ref{sec:detailedBlitza} brings the details and Algorithm~\ref{alg:urbTOV} exemplifies the indication handling.

\Section{System settings}
\label{sec:sys}
We consider an asynchronous message-passing system that has no guarantees on the communication delay. Moreover, there is no notion of global (or universal) clocks and the algorithm cannot explicitly access the local clock (or timeout mechanisms). The system consists of a set, $\sP=\{p_0,\ldots, p_{n-1}\}$, of $n$ crash-prone nodes (or processors) with unique identifiers. Due to an impossibility~\cite[Chapter 3.2]{DBLP:books/mit/Dolev2000}, we assume that any pair of nodes $p_i,p_j \in \sP$ have access to a bidirectional communication channel, $\mathit{channel}_{j,i}$, that, at any time, has at most $\capacity \in \N$ packets on transit from $p_j$ to $p_i$. 

%
%
In the \emph{interleaving model}~\cite{DBLP:books/mit/Dolev2000}, the node's program is a sequence of \emph{(atomic) steps}. Each step starts with an internal computation and finishes with a single communication operation, \ie a message $send$ or $receive$. The \emph{state}, $s_i$, of node $p_i \in \sP$ includes all of $p_i$'s variables and $\mathit{channel}_{j,i}$. The term \emph{system state} (or configuration) refers to the tuple $c = (s_1, s_2, \cdots,  s_n)$. We define an \emph{execution (or run)} $R={c[0],a[0],c[1],a[1],\ldots}$ as an alternating sequence of system states $c[x]$ and (atomic) steps $a[x]$, such that each $c[x+1]$, except for the starting one, $c[0]$, is obtained from $c[x]$ by the execution of step $a[x]$ that some processor takes.  The set of \emph{legal executions} ($LE$) refers to all the executions in which the requirements of the task $T$ hold.


\Subsection{The fault model and self-stabilization}
Failures are environment steps rather than algorithm steps.

\Subsubsection{Benign failures}
\label{sec:benignFailures}
When the occurrence of a failure cannot cause the system execution to lose legality, \ie to leave $LE$, we refer to that failure as a benign one. 
%
%
The system is prone to \emph{crash failures}, in which nodes stop taking steps forever. We assume that at most $t<  n/2 $ node may crash. We denote by $\mathit{Correct}$ the set of indices of processors that never crash. 
%
%
%
We consider solutions that are oriented towards asynchronous message-passing systems and thus they are oblivious to the time in which the packets arrive and depart. Also, the communication channels are prone to packet failures, such as omission, duplication, reordering. However, if $p_i$ sends a message infinitely often to $p_j$, node $p_j$ receives that message infinitely often. We refer to the latter as the \emph{fair communication} assumption. We assume that any message can reside in a communication channel only for a finite period (before it is delivered or lost). The length of that period is unbounded since we assume no bound on transmission delays. In other words, our communication model formally excludes messages that it takes an infinite time to deliver or loss them; but to say that it takes an infinite time to deliver a given message means that this message is lost.

\Subsubsection{Arbitrary transient-faults}
We consider any violation of the assumptions according to which the system was designed to operate. We refer to these violations and deviations as \emph{arbitrary transient-faults} and assume that they can corrupt the system state arbitrarily (while keeping the program code intact). The occurrence of an arbitrary transient-fault is rare. Thus, our model assumes that the last arbitrary transient-fault occurs before the system execution starts~\cite{DBLP:books/mit/Dolev2000}. Also, it leaves the system to start in an arbitrary state.

\Subsubsection{Dijkstra's self-stabilization criterion}
\label{sec:Dijkstra}
An algorithm is \textit{self-stabilizing} with respect to the task of $LE$, when every (unbounded) execution $R$ of the algorithm reaches within a finite period a suffix $R_{legal} \in LE$ that is legal. That is, Dijkstra~\cite{DBLP:journals/cacm/Dijkstra74} requires that $\forall R:\exists R': R=R' \circ R_{legal} \land R_{legal} \in LE \land |R'| \in \mathbb{Z}^+$, where the operator $\circ$ denotes that $R=R' \circ R''$ concatenates $R'$ with $R''$. The complexity measure of self-stabilizing systems, called \emph{stabilization time}, is the time it takes the system to recover after the occurrence of the last transient-fault, \ie $|R'|$. The studied and proposed solutions allow nodes to interact and share information via binary consensus objects and uniform reliable broadcast (URB). Thus, we measure the stabilization time as the number of accesses to these primitives plus the number of URB accesses.      



\Section{Background: Non-self-stabilizing Non-blocking Multivalued Consensus}
\label{sec:back}
We review in sections~\ref{sec:nonSeflStab} and~\ref{sec:seflStabRes} a non-self-stabilizing non-blocking algorithm for multivalued consensus by Most{\'{e}}faoui, Raynal, and Tronel~\cite{DBLP:journals/ipl/MostefaouiRT00}, which uses an unbounded number of binary consensus objects.

\begin{algorithm}[t!]
	\begin{\algSize}
		%
		
		\smallskip
		
		\noindent \textbf{local variables:}\\
		%
		$proposals[0,.,n\text{-}1]$ \tcc*{array of the received proposals}\label{ln:proposalsBotsV}
		$k$ \tcc*{the round counter}\label{ln:kZero}
		$BC[0,.,n\text{-}1]$ \tcc*{binary consensus objects (unbounded list)}\label{ln:BCZero}

		\smallskip
		
		\textbf{operation} $\mathsf{propose}(v)$ \label{ln:proposeV}\Begin{
			$(proposals,BC) \gets ([\bot, \ldots ,\bot],[\bot, \ldots ,\bot])$\label{ln:kProposalGetsZero}\;
			$\mathsf{urbBroadcast}~\mathrm{PROPOSAL}(v)$\label{ln:urbProposalV}\;
			\While{$(k\gets 0; \true; k \gets k+1)$\label{ln:whileLoop}}{
				\If{$BC[k].\mathsf{binPropose}( (proposals[k \bmod n] \neq \bot))$\label{ln:resTrueIf}}{{$\mathbf{wait}(proposals[k \bmod n] \neq \bot)$\label{ln:waitMe}\; \Return{$(proposals[k \bmod n])$}\label{ln:kGetsKplusOne}\;}}
			}
		}
		
		\smallskip
		
		\textbf{upon} $\mathsf{urbDelivered}~\mathrm{PROPOSAL}(v)$ \textbf{from} $p_j$ \textbf{do} \{$proposals[j] \gets v$;\label{ln:proposalsJgetsVdelivered}\}		
		
		\smallskip
		
		\caption{\label{alg:consensusNon}Non-self-stabilizing non-blocking multivalued consensus using an unbounded number of binary consensus instances; code for $p_i$}
	\end{\algSize}
\end{algorithm}

\Subsection{Algorithm~\ref{alg:consensusNon}: non-self-stabilizing multivalued consensus}
\label{sec:nonSeflStab}
The non-self-stabilizing solution in Algorithm~\ref{alg:consensusNon} is the basis for its self-stabilizing variation in Algorithm~\ref{alg:consensus}. For the sake of a simple presentation, the line numbers of Algorithm~\ref{alg:consensus} continues the ones of Algorithm~\ref{alg:consensusNon}.
The operation $\mathsf{propose}(v)$ (line~\ref{ln:proposeV}) invokes an instance of a multivalued consensus object. Algorithm~\ref{alg:consensusNon} uses a uniform reliable broadcast (URB)~\cite{DBLP:books/sp/Raynal18} for letting any $p_i \in \sP$ disseminate its proposed value $v_i$ (line~\ref{ln:urbProposalV}). Each $p_j \in \sP$ that delivers this proposal, stores this value in $proposal_j[i]$. Also, $p_k \in \sP$ can concurrently broadcast its proposal, $v_k$, which $p_j$ stores in $proposal_j[k]$. Therefore, Algorithm~\ref{alg:consensusNon} needs to decide which entry in $proposal_j[]$ the $\mathsf{propose}(v)$ operation should return. This decision is coordinated via an unbounded global array $BC[0], BC[1], \ldots$, of binary consensus objects.

Algorithm~\ref{alg:consensusNon} starts by URB-broadcasting $p_i$'s proposed value, $v_i$ (line~\ref{ln:urbProposalV}). This broadcast assures that all correct nodes receive identical sets of messages (Section~\ref{sec:URB}). Also, the set of delivered messages must include every message URB-broadcast by any correct node. The arrival of $\mathrm{PROPOSAL}(v)$ from $p_j$, informs $p_i$ about $p_j$'s proposal, and thus, $p_i$ stores $v_j$ in $proposals_i[j]$ (line~\ref{ln:proposalsJgetsVdelivered}).

Following the proposal broadcast, Algorithm~\ref{alg:consensusNon} proceeds in asynchronous rounds. The variable $k$ stores the round counter (lines~\ref{ln:kZero} and~\ref{ln:whileLoop}).
	%
Once $p_i$ decides, it leaves the loop by returning the value of $proposals_i[x_i]$ (line~\ref{ln:kGetsKplusOne}), where $x_i=k_i \bmod n$. In other words, $x_i \in \{0, \ldots , n \mathit{-} 1\}$ is the identifier of the node that has broadcast the proposal stored in $proposals_i[x_i]$.

As mentioned, the selection of $x_i$ is facilitated via the unbounded array, $BC[]$, of binary consensus objects. Since all correct nodes eventually receive the same set of broadcasts, $p_i$ proposes $proposals_i[x] \neq\bot$ to the $k_i$-th object, $BC[k_i]$ (line~\ref{ln:resTrueIf}). \Ie $p_i$ proposes $\true$ on the $k_i$-th round if, and only if, it received $p_{x_i}$'s proposal.

Algorithm~\ref{alg:consensusNon} continues to the next round whenever $BC[k_i]$ decides $\false$. Otherwise, $p_i$ decides the value, $proposals_i[x_i]$, proposed by $p_{x_i}$. Due to asynchrony, $p_i$ might need to wait until $p_{x_i}$'s broadcast was URB-delivers (line~\ref{ln:waitMe}). However, if any node proposed to decide $v_{x_i}$, it must be the case that $proposals_i[x_i]$ was delivered to the node that has proposed $\true$ at $BC[k_i]$. Therefore, eventually, $p_i$ is guaranteed to URB-deliver $v_{x_i}$ and stores it at $proposals_i[x_i]$. For this reason, Algorithm~\ref{alg:consensusNon} does not block forever in line~\ref{ln:waitMe} and the decided value is eventually returned in line~\ref{ln:kGetsKplusOne}.

\remove{

Following the proposal broadcast, Algorithm~\ref{alg:consensusNon} proceeds in asynchronous rounds. The variable $k$ stores the round counter (line~\ref{ln:kZero}), such that round $k'$ considers the proposal stored in $proposals[k'\bmod n]$ (lines~\ref{ln:proposeV} to~\ref{ln:kGetsKplusOne}). 
Note that this happens after the broadcasting of $p_i$'s proposal, which implies the instant self-delivery of $p_i$'s proposal and that $p_i$ initializes the variables $proposals[]$ and $k$ before repeatedly attempting to agree whether $proposals_i[k' \bmod n]$ includes a non-$\bot$ value (line~\ref{ln:letBinPropProposal}). 
This agreement is facilitated by an unbounded list of binary consensus objects, $BC[]$. On round $k'$, the value in $BC_i[k']$ stores the decision eventually. Since these rounds are not synchronized, it can be the case that it takes $k'\geq n$ rounds until the binary consensus object $BC[k']$ returns the value $\true$. Therefore, the module function is used when retrieving proposals, \ie $proposals_i[k' \bmod n]$. This sequence of invocations can continue as the for loop (lines~\ref{ln:proposeV} to~\ref{ln:kGetsKplusOne}) iterates. The loop ends when the if-statement condition in line~\ref{ln:resTrueIf} holds, \ie $BC_i[k'].\mathsf{binPropose}()$ returns $\true$ (line~\ref{ln:resBCkBin}). In this case, $p_i$ waits until $proposals_i[k' \bmod n]$ indeed stores a value that is non-$\bot$ (this is needed due to asynchrony). Then, the $\mathsf{propose}()$ operation returns $proposals_i[k' \bmod n]$.

} 

\Subsection{Executing Algorithm~\ref{alg:consensusNon} in the presence of transient-faults}
\label{sec:seflStabRes}
Before describing Algorithm~\ref{alg:consensus}, we review the main challenges that one faces when transferring Algorithm~\ref{alg:consensusNon} to an algorithm that can recover after the concurrence of transient-faults. 

\Subsubsection{Use of an unbounded number of binary objects}
\label{sec:unbBC}
Self-stabilizing systems can only use a bounded amount of memory~\cite{DBLP:books/mit/Dolev2000}. This is because, in practice, computer systems can use only a finite amount of memory. However, a single transient-fault can set every counter (or data-structure) to its maximum value (respectively, exhaust the memory capacity of the data-structure).


\Subsubsection{Corrupted round number counter}
\label{sec:coRndNum}
In the context of self-stabilization, one cannot simply rely on counter $k$ (line~\ref{ln:kZero}) to count the number of asynchronous rounds. This is because a single transient-fault can set the value of $k$ to zero. It can also alter the state of every $BC[k]_{k \in \{0,\ldots, z\} \land z \in \mathbb{Z}^+}$, such that a call to $\mathsf{propose}()$ returns $\false$, where $z$ can be practically infinite, say $z=2^{64}-1$. In this case, the system will have to iterate for $2^{64}$ times before a fresh binary consensus object is reached. 



\Subsubsection{Corrupted program counter}
\label{sec:coProgCnt}
%
%
A transient-fault can set the program counter of every $p_j \in \sP$ to skip over the broadcast in line~\ref{ln:urbProposalV} and to point to line~\ref{ln:whileLoop}. 
If this happens, then validity or termination can be violated. Therefore, there is a need to repeat the transmission of $v_i$ in order to make sure that at least one proposal is known to all correct processors.


\Subsubsection{A corrupted array of binary objects}
\label{sec:unbBCA}
Transient faults can corrupt binary objects in the array $BC[]$. Specifically, since the array $BC[]$ should include only a bounded number of binary consensus objects, a transient-fault can change the state of all objects in $BC[]$ to encode `decide $\false$'. In this case, Algorithm~\ref{alg:consensusNon} cannot finish the multivalued consensus. 

\begin{figure}[t!]
	\begin{framed}
		\begin{\algSize}
			
			\smallskip
			
			\textbf{(a)} Upon $\mathsf{propose}(v)$, uniform reliable broadcast $\langle v \rangle$.
			
			\textbf{(b)} By URB-termination, eventually, there is $p_j \in \sP$ and round $k'$, such that $p_j$'s message arrived at all non-faulty processors, \ie $\forall  \ell \in \mathit{Correct} \implies proposals_\ell[j]\neq \bot$.
			
			\textbf{(c)} For $k \in\{0,1,2,\ldots\}$, $p_i$ invokes $BC[k].\mathsf{binPropose}(proposals[k \bmod n] \neq \bot)$.
			
			\textbf{(d)} By BC-termination and stage (b), eventually, the $k_{\min}$-th binary consensus objects is the first to decide $\true$ while all $x$-th objects decide $\false$, where $x \in \{0,1,2,k_{\min}\mathit{-}1\}$. 
			
			\textbf{(e)} Due to URB-termination, eventually, $proposals[k_{\min} \bmod n]$ includes a non-$\bot$ value. 
			
			\textbf{(f)} Then, return $proposals[k_{\min} \bmod n]$ as the decided value.
			
			\smallskip
		\end{\algSize}
	\end{framed}
	
	\caption{\label{alg:HconsensusNon}High-level stages in the execution of Algorithm~\ref{alg:consensusNon}; code for $p_i$}
\end{figure}

\begin{figure}[h!]
	\begin{framed}
		\begin{\algSize}
			
			\smallskip
			
			\textbf{(a)}  Upon $\mathsf{propose}(v)$, uniform reliable broadcast $\langle v \rangle$.
			
			\textbf{(b)} \fbox{Wait until $\exist()$ says that $\langle v \rangle$ arrived at all non-faulty processors.}
			
			\textbf{(c)} For \fbox{$k \in \{0,\ldots,n\mathit{-}1\}$}, $p_i$ invokes $BC[k].\mathsf{binPropose}(proposals[k \bmod n] \neq \bot)$. 
			
			\textbf{(d)} By BC-termination and stage (b), eventually, the $k_{\min}$-th binary consensus objects is the first to decide $\true$ while all $x$-th objects decide $\false$, where $x \in \{0,1,2,k_{\min}\mathit{-}1\}$. 
			
			\textbf{(e)} Due to URB-termination, eventually, $proposals[k_{\min} \bmod n]$ includes a non-$\bot$ value. 
			
			\textbf{(f)} Then, return $proposals[k_{\min} \bmod n]$ as the decided value.
			
			\smallskip
		\end{\algSize}
	\end{framed}
	
	\caption{\label{alg:Hconsensus}A bounded alternative to Figure~\ref{alg:HconsensusNon}; code for $p_i$}
	
\end{figure}

\Section{The Proposed Solution: Self-stabilizing Wait-free Multivalued Consensus}
\label{sec:seflStab}
%
%
This section presents a new self-stabilizing algorithm for multivalued consensus that is wait-free and uses $n$ binary consensus objects and $n$ self-stabilizing uniform reliable broadcasts (URBs)~\cite{selfStabURB}. The correctness proof appears in Section~\ref{sec:corr}.

\Subsection{The algorithm idea}
We sketch the key notions that are needed for Algorithm~~\ref{alg:consensus} by addressing the challenges raised in Section~\ref{sec:seflStabRes}.

\Subsubsection{Using a bounded number of binary objects}
We explain how Algorithm~\ref{alg:consensus} can use only at most $n$ binary consensus objects. Figure~\ref{alg:HconsensusNon} is a high-level description of Algorithm~\ref{alg:HconsensusNon}'s execution and Figure~\ref{alg:Hconsensus} shows how this process can be revised. The key differences between figures~\ref{alg:HconsensusNon} and~\ref{alg:Hconsensus} appear in the \fbox{boxed text} of Figure~\ref{alg:Hconsensus}. Specifically, Figure~\ref{alg:Hconsensus} waits until $p_i$'s broadcast has terminated in line (b). At that point in time, $p_i$ knows that all non-faulty processors have received its message. Only then does $p_i$ allow itself to propose values via the array of binary objects. This means that no processor starts proposing any binary value before there is at least one index $k \in \{0,\ldots, n\mathit{-}1\}$ for which $proposals_j[k] \neq \bot$, where $p_j$ is any node that has not failed. This means that, regardless of who is going to invoke the $k$-th binary consensus object, only the value $\true$ can be proposed. For this reason, there is no need to use more than $n$ binary consensus values until at least one of them decides $\true$, cf. line (c) in Figure~\ref{alg:Hconsensus}.

\Subsubsection{Dealing with corrupted round number counter}
Using the object values in $BC[]$, Algorithm~\ref{alg:consensus} calculates $k()$, which returns the current round number. This way, a transient-fault cannot create inconsistencies between $k()$'s value and $BC[]$.

In detail, for an active multivalued consensus object $O$, \ie $O \neq \bot$, we say that the binary consensus object $O.BC[k]$ is active when $O.BC[k]\neq \bot$. Algorithm~\ref{alg:consensus} calculates $\mathsf{k}()$ (line~\ref{ln:KsMBC}) by counting the number of active binary consensus objects that have terminated and the decided value is $\false$. We restrict this counting to consider only the entries $BC[k]$, such that $k=0$ or $\forall k'<k:BC[k']$ is an active binary consensus objects that have terminated and the decided $\false$. This is defined by the set $\mathsf{K} =  (\{k \in S(n\text{-}1):O.BC[k]\neq \bot$ $ \land O.BC[k].\done(k) =\false \})$, where $S(x) =\{0,\ldots, x\}$ is the set of all integers between zero and $x$. This way, the value of $\mathsf{k}()$ is simply $\max(\{\{\mathit{-}1\} \cup\{x \in S(n\mathit{-}1): (S(x)\cap \mathsf{K})= S(x) \})$. Note that the value of $\mathit{-}1$ is used to indicate that there are no active binary objects in $BC[]$ that have terminated with a decided value of $\false$, \ie $\mathsf{K}=\emptyset$.

\Subsubsection{Dealing with a corrupted program counter}
%
%
\label{sec:DealingArrray}
As explained in Section~\ref{sec:coProgCnt}, there is a need to repeat the transmission of $v_i$ in order to make sure that at least one proposal is known to all correct processors. Specifically, after $\mathsf{propose}_i(v_i)$'s invocation, $p_i \in \sP$ need to store $v_i$ and broadcasts $v_i$ repeatedly due to a well-known impossibility~\cite[Chapter 2.3]{DBLP:books/mit/Dolev2000}. Note that there is an easy way to trade the broadcast repetition rate with the recovery speed from transient-faults. Also, once the first broadcast has terminated, all correct processors $p_i \in \sP$ are ready to decide by proposing $\mathsf{binPropose}_i(k,proposals_i[k]\neq \bot)$ for any $p_k \in \sP$, see steps (b) and (c) in Figure~\ref{alg:Hconsensus}.

\Subsubsection{Dealing with a corrupted array of binary objects}
Algorithm~\ref{alg:consensus} uses only $n$ binary consensus objects. Due to the challenge in Section~\ref{sec:unbBCA}, we explain how to deal with the case in which a transient-fault changes the state of all objects in $BC[]$ to encode `decide $\false$'. In this case, the algorithm cannot satisfy the requirements of the multivalued consensus task (Definition~\ref{def:consensus}). Therefore, our solution identifies such situations and informs the invoking algorithm via the return of the \emph{transient error} symbol $\blitza$.

\Subsection{Algorithm description}

\begin{algorithm*}[t!]
	\begin{\algSize}	
		
		
				\noindent \textbf{variables:}\texttt{ /* initialization is optional in the context of self-stabilization */} 
		
		$v$ \tcc*{local decision estimates}
		$proposals[0,.,n\text{-}1]$ \tcc*{array of arriving proposals}
		$BC[0,.,n\text{-}1]$ \tcc*{array of $n$ binary consensus objects}
		$\txD$ \tcc*{URB transmission descriptor for decision sharing}
		$oneTerm$ \tcc*{true once at least one broadcast termination occured}

		\smallskip
		
		
		\textbf{macro} $\mathsf{k}() = \max(\{\{\text{-}1\} \cup\{x \in S(n\text{-}1): (S(x)\cap \mathsf{K})= S(x) \})$\label{ln:curKs}: \textbf{where} $S(x)=$ $\{0,\ldots, x\}$ \textbf{and} $\mathsf{K} =  (\{k \in S(n\text{-}1):O.BC[k]\neq \bot$ $ \land O.BC[k].\done(k) =\false \})$\label{ln:KsMBC} \tcc*{$\mathsf{k}()$ is the max consecutive $BC[]$ entry index with the decision $\false$}
		
		\smallskip
		
		
		\textbf{operation} $\mathsf{propose}(\remove{s,}v)$ \textbf{do} \{\lIf{\label{ln:testVbot}$v \neq \bot \land O=\bot$\label{ln:CSsMdCSlorCS}}{$O.(v,proposals,BC,\txD,oneTerm) \gets (v,[\bot, \ldots ,\bot],[\bot, \ldots ,\bot],\bot,\false)$\label{ln:vBasic}\}}
		
		\smallskip

\textbf{operation} $\done()$ \label{ln:doneConsensus}\Begin{
	\lIf{$O=\bot$}{\Return{$\bot$}\label{ln:basicNeqTestBot}}
	\lElseIf{$O.v = \bot \lor k\geq n -1$\label{ln:blitza}}{\Return{$\blitza$} \textbf{where} $k=\mathsf{k}()$}
	\lElseIf{$BC[k+1]=\bot \lor BC[k+1].\done(k+1)\neq\true$\label{ln:OKbmodN}}{\Return{$\bot$}} 
	\lElseIf{$x=\bot$}{\Return{$\blitza$} \textbf{else} \Return{$x$} \textbf{where} $x=O.proposals[k+1]$\label{ln:returnedDecidedValue}}
}
		\smallskip
		
		
		\textbf{do forever} {\ForEach{$O\neq \bot$ \textbf{\emph{with}} $O$'s \textbf{\emph{fields}} \remove{\mathit{seq},} $v$, $proposals$, $BC$, \textbf{\emph{and}} $\txD$\label{ln:ellIn0M1}}
			{{
					
					\If{$(v \neq \bot \land (\txD=\bot \lor  \exist(\txD))$\label{ln:txDBotLandProp}}{
						$oneTerm \gets oneTerm  \lor (\txD \neq\bot \land \exist(\txD))$\label{ln:vURBfix}\;
						$\txD \gets \mathsf{urbBroadcast}~\mathrm{PROPOSAL}(\remove{\mathit{seq},}v)$\label{ln:vURB}
					}
					
					\tcc{use either lines~\ref{ln:proposalsIneqBotA} to~\ref{ln:binProposeSeqProposalsA} or lines~\ref{ln:proposalsIneqBotB} to~\ref{ln:binProposeSeqProposalsB}}
					\If{$oneTerm \land \mathsf{k}< n\text{-}1\land BC[\mathsf{k}\mathit{+}1]$$=$$\bot \land (\mathsf{k}$$=$$\text{-}1 \lor BC[\mathsf{k}].\done(\mathsf{k}) \neq \bot)$\label{ln:proposalsIneqBotA}}{$\mathsf{binPropose}(\remove{\mathit{seq},}\mathsf{k}\text{+}1,proposals[\mathsf{k}\text{+}1] \neq \bot)$\label{ln:binProposeSeqProposalsA} \textbf{where} $k=\mathsf{k}()$}
										
					\If(\texttt{/* invoke BC objects concurrently */}){\fbox{$oneTerm \land \exists \ell :BC[\ell]=\bot$\label{ln:proposalsIneqBotB}}}{\fbox{\textbf{for each} $k \in \{0,\ldots, n\mathit{-}1\}:BC[k]=\bot$ \textbf{do} $\mathsf{binPropose}(k,proposals[k\mathit{+}1] \neq \bot)$\label{ln:binProposeSeqProposalsB}}}				
			}}}

		\smallskip
		
		
		%
		
		\textbf{upon} $\mathrm{PROPOSAL}(\mathit{vJ})$ $\mathsf{urbDelivered}$ \textbf{from} $p_j$ \Begin{
			
			\If{$ \mathit{vJ} \neq \bot$\label{ln:testSjVjNeqBot}}{
				\lIf{$O\neq \bot \land O.proposals[j]=\bot$}{$O.proposals[j] \gets \mathit{vJ}$\label{ln:proposalsGetsVj}}
				\lElseIf{$O=\bot$}{$(O.(\remove{\remove{\mathit{seq},}}v,proposals,BC,\txD)$, $O.proposals[j])$$\gets$$((\mathit{sJ},\mathit{vJ},[\bot, \ldots ,\bot], [\bot, \ldots ,\bot],\bot),\mathit{vJ})$\label{ln:OgetsVj}}
			}
		}

		
		\smallskip
		
		\caption{\label{alg:consensus}Self-stabilizing\reduce{ non-blocking} multivalued consensus; $p_i$'s code}
	\end{\algSize}
\end{algorithm*}

\Subsubsection{The $\mathsf{propose}(v)$ operation and variables}
The operation $\mathsf{propose}(\remove{s,}v)$ activates a multivalued object by initializing its fields (line~\ref{ln:testVbot}). These are \remove{the sequence numbers, $\mathit{seq}$, }the proposed value, $v$, the array, $proposals[]$, of received proposals, where $proposals[j]$ stores the value received from $p_j \in \sP$. Moreover, $BC[]$ is the array of binary consensus objects, where the active object $BC[j]$ determines whether the value in $proposals[j]$ should be the decided value. Also, $\txD$ is the transmission descriptor (initialized with $\bot$), and $oneTerm$ is a boolean that indicates that at least one transmission has completed, which is initialized with $\false$. Note that only $v$ has its (immutable) value initialized in line~\ref{ln:vBasic} to its final value. The other fields are initialized to $\bot$ or an array of $\bot$ values; their values can change later on.


\Subsubsection{The $\done()$ operation}
\label{sec:detailedBlitza}
Algorithm~\ref{alg:consensus} allows retrieving the decided value via $\done(\remove{s})$ (line~\ref{ln:doneConsensus}). As long as the multivalued consensus object is not active (line~\ref{ln:basicNeqTestBot}), or there is no decision yet (line~\ref{ln:OKbmodN}), the operation returns $\bot$. As explained in Section~\ref{sec:DealingArrray}, Algorithm~\ref{alg:consensus} might enter an error state. In this case, $\done()$ returns $\blitza$ (lines~\ref{ln:blitza} and~\ref{ln:returnedDecidedValue}). The only case that is left (the else clause of line~\ref{ln:returnedDecidedValue}) is when there is a binary consensus object $O.BC[k]$ and a matching $O.proposals[k] \neq \bot$, where $k=k()$. Here, due to the definition of $k()$ (line~\ref{ln:curKs}), for any $k'\in \{0,\ldots, k\text{-}1\}$ the decided value of $O.BC[k']$ is $\false$ and $O.BC[k]$ decides $\true$. Thus,  $\done_i()$ returns the value of $O.proposals[k]$. 


\Subsubsection{The do-forever loop}
As explained above, Algorithm~\ref{alg:consensus} has to make sure that the proposed value, $v$, arrives at all processors and records in $oneTerm$ the fact that at least once transmission has arrived.
%
%
To that end, in line~\ref{ln:txDBotLandProp}, $p_i$ tests the predicate $(\txD \neq \bot \land \exist(\txD))$ and makes sure that the transmission descriptor, $\txD$, refers to an active broadcast, \ie $\txD$ stores a descriptor that has not terminated (cf. $\exist()$'s definition in Section~\ref{sec:URB}). In detail, whenever $x.\txD \neq \bot$ holds, $\exist_i(\txD)$ holds eventually (URB-termination). Thus, the if-statement condition in line~\ref{ln:txDBotLandProp} holds eventually and $p_i$ URB-broadcast $\mathrm{PROPOSAL}(v)$ (line~\ref{ln:vURB}) after checking that $v\neq \bot$ (line~\ref{ln:txDBotLandProp}). Note that $p_i$ records the fact that at least one transmission was completed by assigning $\true$ to $oneTerm$ (line~\ref{ln:vURBfix}). 

Upon the URB-delivery of $p_i$'s $\mathrm{PROPOSAL}(\mathit{vJ})$ at $p_j \in \sP$, \ems{processor $p_j$ considers} the following two cases. If $O$ is an active object, $p_j$ merely checks whether $O.proposals[i]$ needs to be updated with $\mathit{vJ}$ (line~\ref{ln:proposalsGetsVj}). Otherwise, $O$ is initialized with $\mathit{vJ}$ as the proposed value (line~\ref{ln:OgetsVj}) similarly to line~\ref{ln:vBasic}.

Going back to the sender side, Algorithm~\ref{alg:consensus} uses either lines~\ref{ln:proposalsIneqBotA} to~\ref{ln:binProposeSeqProposalsA}, which sequentially access the array, $BC[]$, of binary consensus objects, or lines~\ref{ln:proposalsIneqBotB} to~\ref{ln:binProposeSeqProposalsB}, which simply access all binary objects concurrently.
In both methods, processor $p_i$ makes sure that at least one broadcast was completed, \ie $oneTerm=\true$ (lines~\ref{ln:vURBfix},~\ref{ln:proposalsIneqBotA} and~\ref{ln:proposalsIneqBotB}). When following the sequential method (lines~\ref{ln:proposalsIneqBotA} to~\ref{ln:binProposeSeqProposalsA}), the aim is to invoke binary consensus by calling $\mathsf{binPropose}(k\text{+}1,proposals[k\text{+}1] \neq \bot)$ (line~\ref{ln:binProposeSeqProposalsA}), where $k=\mathsf{k}_i()$. This can only happen when the $(k\text{+}1)$-th object in $BC[]$ is not active, \ie $BC[k\text{+}1] = \bot$ and $BC[k\text{+}1]$ is either the first in $BC[]$, \ie $k=\text{-}1$ or $BC[k]$ has terminated, \ie $BC[k].\done_i(k) \neq\bot$ (line~\ref{ln:proposalsIneqBotA}).


\ems{The advantage of the sequential access method over the concurrent one is that it is more conservative with respect to the number of consensus objects that are being used since once the decision is $\true$, there is no need to use more objects. The concurrent access method, marked in the boxed lines, encourages to piggyback of the messages related to binary concurrent objects. This is most relevant when every message (of binary consensus) can carry the data-loads of $n$ proposals. In this case, the concurrent access method is both simpler and faster than the sequential one.}             

\Section{Correctness of Algorithm~\ref{alg:consensus}}
\label{sec:corr}
Theorems~\ref{thm:recoveryMultivalued} and~\ref{thm:closureMultivalued} show that Algorithm~\ref{alg:consensus} implements a self-stabilizing multivalued consensus. Definition~\ref{def:consistent} is used by Theorem~\ref{thm:recoveryMultivalued}. As explained in Section~\ref{sec:org}, for the sake of a simple presentation, we make the following assumptions. Let $R$ be an Algorithm~\ref{alg:consensus}'s execution, $p_i \in \sP$, and $O_i$ a multivalued consensus object.

\begin{definition}[Consistent multivalued consensus object]
	\label{def:consistent}
	Let $R$ be an Algorithm~\ref{alg:consensus}'s execution and $O_i$ a multivalued consensus object, where $p_i \in \sP$. Suppose either (i) $O_i=\bot$ is inactive or that (ii) $O_i \neq \bot$ is active, $O_i.v \neq \bot \land (k< n -1) \land ((BC[k\mathit{+}1] = \bot \lor BC[k\mathit{+}1].\done(k\mathit{+}1) = \bot \lor (BC[k\mathit{+}1].\done(k\mathit{+}1)=\true \land O_i.proposals[k\mathit{+}1] \neq \bot)))$, where $k=\mathsf{k}_i()$. In either case, we say that $O_i$ is consistent in $c$.
\end{definition}

Theorem~\ref{thm:recoveryMultivalued} shows recovery from arbitrary transient-faults.

\begin{theorem}[Convergence] 
	\label{thm:recoveryMultivalued}
	Let $R$ be an Algorithm~\ref{alg:consensus}'s execution. Suppose that there exists a correct processor $p_j \in \sP: j \in \mathit{Correct}$, such that throughout $R$ it holds that $O_j \neq \bot$ is an active multivalued consensus object. Moreover, suppose that any correct processor $p_i \in \sP: i \in \mathit{Correct}$ calls $\done_i()$ infinitely often in $R$. Within $n$ invocations of binary consensus, (i) the system reaches a state $c \in R$ after which $\done_i() \neq \bot$ holds. Specifically, (ii) $O_i$ is either consistent (Definition of~\ref{def:consistent}) or eventually reports the occurrence of a transient-fault, \ie $\done_i() = \blitza$.
\end{theorem}  
\renewcommand{\thmcnt}{\ref{thm:recoveryMultivalued}}
\begin{theoremProof}
Lemmas~\ref{thm:doneNeqBot} and~\ref{thm:eitherConsistent} implies the proof.
	\begin{lemma}
		\label{thm:doneNeqBot}
Invariant (i) holds, \ie  $\done_i() \neq \bot$ holds in $c$.  
	\end{lemma}
	\renewcommand{\lemcnt}{\ref{thm:eitherConsistent}}
	\begin{lemmaProof}
	Suppose, towards a contradiction, that $c$ does not exist. Specifically, let $R'$ be the longest prefix of $R$ that includes no more than $n$ invocations of binary consensus. The proof of Invariant (i) needs to show that the system reaches a contradiction by showing that $c \in R'$. To that end, arguments (1) to (3), as well as claims~\ref{thm:stopIfandIF} to~\ref{thm:stopIfandIFA}, show the needed contradiction.

	\F
	Argument (1) implies that it is enough to show that the if-statement in line~\ref{ln:OKbmodN} cannot hold eventually. 
	
	\textbf{Argument (1)} \emph{The if-statement conditions in lines~\ref{ln:basicNeqTestBot},~\ref{ln:blitza}, and~\ref{ln:returnedDecidedValue} do not hold for $p_j$ throughout $R$.~~}
	By the theorem assumption that $O_j \neq \bot$ is an active multivalued consensus object throughout $R$, we know that the if-statement condition in line~\ref{ln:basicNeqTestBot} cannot hold.
	Moreover, by the assumption that $c$ does not exist, we know that the if-statement conditions in lines~\ref{ln:blitza} and~\ref{ln:returnedDecidedValue} do not hold for any (correct) $p_i$ throughout $R$. 
	
	\F
	\textbf{Argument (2)} \emph{The invariant $O_j.v \neq \bot$ holds throughout $R$.~~}
	%
	Since the if-statement condition in line~\ref{ln:blitza} does not hold, $O_j.v \neq \bot$ holds in $R$'s starting system state. Moreover, only lines~\ref{ln:vBasic},~\ref{ln:proposalsGetsVj}, and~\ref{ln:OgetsVj} change the value of $O_j.v$ but this happens only after testing that the assigned value is not $\bot$ (lines~\ref{ln:testVbot} and~\ref{ln:testSjVjNeqBot}). 

	\F	
	\textbf{Argument (3)} \emph{$R$ has a suffix in which all correct processors $p_i \in \sP$ are active.~~}
	Since $p_j$ is active and $O_j.v \neq \bot$ holds throughout $R$, the if-statement condition in line~\ref{ln:txDBotLandProp} holds eventually since either $\txD=\bot$ or $\exist(\txD)$ holds eventually due to the URB-termination property. By line~\ref{ln:vURB}, $p_j$ broadcasts the $\langle v \rangle$ message to all correct processors $p_i$. By the URB-termination property, $p_i$ receives $\langle v \rangle$ and by lines~\ref{ln:proposalsGetsVj} to~\ref{ln:OgetsVj}, processor $p_i$ is active.
	
	\F
	
	\textbf{Argument (4)} \emph{$\forall i,j \in \Correct: O_i.proposals[j] \neq \bot \land O_i.\txD \neq \bot \land O_i.oneTerm = \true$ holds eventually.~} 
	By URB-termination, $\exist(O_i.\txD)$ holds eventually. Once that happens, the if-statement condition in line~\ref{ln:txDBotLandProp} holds (due to arguments (2) and (3)) and $O_i.\txD=\bot$ cannot hold (line~\ref{ln:vURB}). By Argument (2), $O_i.v \neq \bot$. Thus, $p_i$ eventually URB-broadcasts $\mathrm{PROPOSAL}(O_i.v)$. Once $p_i$ self-delivers this message, line~\ref{ln:proposalsGetsVj} assigns $v$ to $O_i.proposals[i]$ due to the assumption that $O_i\neq\bot$ throughout $R$. We can now repeat the reasoning that $\exist(O_i.\txD)$ holds eventually and thus the if-statement condition in line~\ref{ln:txDBotLandProp} hold. Thus, $O_i.oneTerm = \false$ does not hold eventually (line~\ref{ln:vURBfix}). By Argument (3), the same holds for $p_j$. Specifically, $p_j$ eventually URB-broadcasts $\mathrm{PROPOSAL}(O_j.v)$. Once $p_i$ URB-delivers this message from $p_j$, $p_i$'s state can possibly change, even in the case that $O_i\neq \bot$, cf. lines~\ref{ln:proposalsGetsVj} and~\ref{ln:OgetsVj}.
	

	\begin{claim}
		\label{thm:stopIfandIF}
		The if-statement condition in lines~\ref{ln:proposalsIneqBotA} and~\ref{ln:proposalsIneqBotB} can only hold at most $n$ times for any $p_i \in \sP: i \in \Correct$.
	\end{claim}
\renewcommand{\clmcnt}{\ref{thm:stopIfandIF}}
\begin{claimProof}
	The if-statement condition in line~\ref{ln:proposalsIneqBotB} can only hold at most once due to line~\ref{ln:binProposeSeqProposalsB}. Thus, the rest of the proof focuses on line~\ref{ln:binProposeSeqProposalsA}. 
	
	By the proof of Argument (4), eventually, the system reaches a state, $c' \in R$, in which $O_i.\txD\neq \bot \land O_i.proposals[j] \neq \bot \land O_i.oneTerm = \true$ holds. Note that the if-statement condition in line~\ref{ln:proposalsIneqBotA} holds whenever $k=\mathit{-}1$. Arguments (5) and (6) assumes that $k>\mathit{-}1$ and consider the cases in which $O_i.BC[k\mathit{+}1] \neq \bot$ holds and does not hold, respectively, where $k=\mathsf{k}()$. Argument (7) shows that the if-statement condition in line~\ref{ln:proposalsIneqBotA} can hold at most $n$ times.  
	
	\F

	\textbf{Argument (5)} \emph{Suppose that $k>\mathit{-}1 \land O_i.BC[k\mathit{+}1] \neq \bot$ holds. Eventually, either $k_i()<n\mathit{-}1$ does not hold or the if-statement condition in line~\ref{ln:proposalsIneqBotA} holds.~} 	
	$BC[k\mathit{+}1].\done_i(k\mathit{+}1) \neq \bot$ holds eventually due to the termination property of binary consensus objects. 
	
	In case $BC[k\mathit{+}1].\done_i(k\mathit{+}1) = \true$, we know that $\done_i() \neq \bot$ holds due to the definition of $k()$ (line~\ref{ln:curKs}). However, this implies a contradiction with the assumption made at the start of this lemma's proof. 
	
	In case $BC[k\mathit{+}1].\done_i(k\mathit{+}1) = \false$ holds, the if-statement condition in line~\ref{ln:proposalsIneqBotA} holds in $c'$ if $k\mathit{+}1 < n\text{-}1$ and $O_i.BC[k\text{+}1]=\bot$. In case the former predicate holds and the latter does not, we can repeat the reasoning above for at most $n$ times until either the former does not hold or both predicates hold. In either case, the proof of the argument is done.
	
	\F

	\textbf{Argument (6)} \emph{Suppose that $k>\mathit{-}1 \land O_i.BC[k\mathit{+}1] = \bot$ holds. Eventually, either $k_i()<n\mathit{-}1$ does not hold or the if-statement condition in line~\ref{ln:proposalsIneqBotA} holds.~} 	
	The if-statement condition in line~\ref{ln:proposalsIneqBotA} holds if $BC[k].\done_i(k) \neq \bot$ holds. 
	Since $k>\mathit{-}1$, the reasoning in the proof of Argument (5), which shows that $BC[k\mathit{+}1].\done_i(k\mathit{+}1) \neq \bot$ holds, can be used for showing that $BC[k].\done_i(k) \neq= \bot$ holds eventually. 
	
	\F
	
	\textbf{Argument (7)} \emph{Within $n$ invocations of $\mathsf{binPropose}_i()$, the if-statement condition in line~\ref{ln:proposalsIneqBotA} does not hold.~} 	
	Suppose that the if-statement condition in line~\ref{ln:proposalsIneqBotA} holds.
	In line~\ref{ln:binProposeSeqProposalsA}, $p_i$ invokes the operation $\mathsf{binPropose}_i(k\text{+}1,O_i.proposals[k\text{+}1] \neq \bot)$ of the $(k\text{+}1)$-\emph{th} binary consensus object. This invocation changes $p_i$'s state, such that $O_i.BC[k\text{+}1]=\bot$ does not hold any longer (because the $\mathsf{binPropose}_i()$ operation initializes the state of $O_i.BC[k\text{+}1]$). Since $BC[]$ has $n$ entries, there could be at most $n$ such invocations until the system reaches $c'' \in R$, after which the if-statement condition in line~\ref{ln:proposalsIneqBotA} cannot hold. 
	\end{claimProof}

	\F

	\begin{claim}
	\label{thm:stopIfandIFA}
	Once the if-statement condition in line~\ref{ln:proposalsIneqBotA} (or~\ref{ln:proposalsIneqBotB}) does not hold, also the if-statement condition in line~\ref{ln:OKbmodN} does not hold.
\end{claim}
\renewcommand{\clmcnt}{\ref{thm:stopIfandIFA}}
\begin{claimProof}
	Since if-statement condition in line~\ref{ln:proposalsIneqBotA} does not hold, we know that $BC_i[k+1]=\bot$ does not hold, see Argument (5) of Claim~\ref{thm:stopIfandIF}. In the case of line~\ref{ln:proposalsIneqBotB}, the same holds in a straight forward manner. By BC-termination, $BC[k+1].\done_i(k+1) \neq \bot$ holds eventually. Since $\forall p_x \in \sP:O_i.BC[x]\neq \bot \land BC[x].\done_i(x)=\false$ implies a contradiction with Argument (1), we know that $BC_i[k+1].\done(k+1)\neq\true$ cannot hold. 
	%
\end{claimProof}
	\end{lemmaProof}
	
\begin{lemma}
	\label{thm:eitherConsistent}
Invariant (ii) holds, \ie  $O_i$ is either consistent or $\done_i() = \blitza$.
\end{lemma}
\renewcommand{\lemcnt}{\ref{thm:eitherConsistent}}
\begin{lemmaProof}
	Recall that the theorem assumes that $O_i$ is an active object throughout $R$. The argument is implied by Definition~\ref{def:consistent} and lines~\ref{ln:basicNeqTestBot} to~\ref{ln:returnedDecidedValue}. 

In detail, line~\ref{ln:basicNeqTestBot} handles the case in which $O_i=\bot$. Suppose that $O_i \neq \bot$ is active, which indicates that an inconsistent was detected. 
Line~\ref{ln:blitza} handles the case in which $O_i.v \neq \bot \land k < n\mathit{-}1$ does not hold by returning $\blitza$, where $k=\mathsf{k}_i()$, which indicates that an inconsistent was detected. 
Line~\ref{ln:OKbmodN} allows the case in which $BC[k\mathit{+}1] = \bot \lor BC[k\mathit{+}1].\done(k\mathit{+}1) = \bot$ (note that the case of $BC[k\mathit{+}1].\done(k\mathit{+}1) = \false$ does not exist due to the definition of $k()$ in line~\ref{ln:curKs}). This case is allowed since it is consistent, see Definition~\ref{def:consistent}. 
Line~\ref{ln:returnedDecidedValue} handles the case in which $(BC[k\mathit{+}1].\done(k\mathit{+}1)=\true \land O_i.proposals[k\mathit{+}1] \neq \bot)$ does not holds by returning $\blitza$, which indicates that an inconsistent was detected.
	\end{lemmaProof}
\end{theoremProof}

\FF

Definition~\ref{def:complete} is used by Theorem~\ref{thm:closureMultivalued}. 

\begin{definition}[Complete execution with respect to $\mathsf{propose}()$ invocations]
	\label{def:complete}
	Let $R$ be an execution of Algorithm~\ref{alg:consensus} that starts in $c \in R$. We say that $c$ is \emph{completely free} of $\mathsf{PROPOSAL}(\bull)$ messages if (i) the communication channels do not include $\mathsf{PROPOSAL}(\bull)$ messages, and (ii) for any non-failing $p_i \in\sP$, there is no active multivalued consensus object $O_i=\bot$ in $c$. 
	Let $c_s \in R$ be the system state that is: (a) completely free of $\mathsf{PROPOSAL}(\bull)$, and (b) it appears in $R$ immediately before a step that includes $p_i$'s invocation of $\mathsf{propose}(\bull)$ (lines~\ref{ln:testVbot}) in which $O_i$ becomes active (rather than due to the arrival of a $\mathsf{PROPOSAL}(\bull)$ message in lines~\ref{ln:testSjVjNeqBot} to~\ref{ln:OgetsVj}). In this case, we say that $p_i$'s invocation is authentic. Suppose that $p_i$ sends a $\mathsf{PROPOSAL}(\bull)$ message after $c_s$. In this case, we say that $\mathsf{PROPOSAL}(\bull)$ is an authentic message transmission. 
	An arrival of $\mathsf{PROPOSAL}(\bull)$ to $p_j \in \sP$ (lines~\ref{ln:testVbot}) is said to be authentic if it is due to an authentic message transmission. Suppose that $p_j$ actives $O_j=CS_j[s]$ (line~\ref{ln:OgetsVj}) due to an authentic arrival (rather than an invocation of the $\mathsf{propose}(\bull)$ operation). In this case, we also say that $p_j$'s invocation is authentic. We complete the definitions of authentic transmissions, arrivals, and invocations by applying the transitive closures of them. 
	%
	%
	Suppose that any invocation in $R$ of $\mathsf{propose}_k(\bull):p_k \in \sP$ is authentic as well as the transmission and reception of $\mathsf{PROPOSAL}(\bull)$ messages from or to $p_k$. In this case, we say that $R$ is authentic. 
\end{definition}

Theorem~\ref{thm:closureMultivalued} shows that Algorithm~\ref{alg:consensus} satisfies the task requirements (Section~\ref{sec:spec}).

\begin{theorem}[Closure] 
	\label{thm:closureMultivalued}
	Let $R$ be an authentic execution of Algorithm~\ref{alg:consensus}. The system demonstrates in $R$ the construction of a multivalued consensus object.
\end{theorem} 
\renewcommand{\thmcnt}{\ref{thm:closureMultivalued}}
\begin{theoremProof}
	Validity holds since only the user input is stored in the field $v$ (line~\ref{ln:vBasic}), which is then URB-broadcast (line~\ref{ln:vURB}), stored in the relevant entry of $proposals$ (lines~\ref{ln:proposalsGetsVj} to~\ref{ln:OgetsVj}), and returned as the decided value (line~\ref{ln:returnedDecidedValue}). Moreover, any value in $v$ can be traced back to an invocation of $\mathsf{propose}(v)$ since $R$ is authentic.
	
\F
	Lemma~\ref{thm:terminationAgreement} demonstrates termination and agreement.

	\begin{lemma}
		\label{thm:terminationAgreement}
		Let $a_i \in R$ be the first step in $R$ that includes an invocation, say, by $p_i \in \sP$ of $\mathsf{propose}_i(v_i)$. Suppose that $v_i \neq \bot$ holds in any system state of $R$. There exists $v \notin \{\bot,\blitza\}$, such that for every correct $p_j \in \sP$ it holds that $\done_j()$ returns $v$ within $n$ invocations of binary consensus.   
	\end{lemma}
\renewcommand{\lemcnt}{\ref{thm:terminationAgreement}}
	\begin{lemmaProof}
		Arguments (1) to (7) imply the proof.
		
		\F
		\textbf{Argument (1)} \emph{$O_i.(v,proposals,BC,\txD,oneTerm)$ $=(v,[\bot, \ldots ,\bot],[\bot, \ldots ,\bot],\bot,\false)$ holds immediately after $a_i$.~~}
		We show that the if-statement condition in line~\ref{ln:CSsMdCSlorCS} holds immediately before $a_i$. Recall the theorem assumption that $v_i \neq \bot$ holds in $R$. By the assumption that $R$ is authentic, we know that $O_i=\bot$ holds immediately before $a_i$. Therefore, $p_i$ assigns $(v_i,[\bot, \ldots ,\bot],[\bot, \ldots ,\bot],\bot,\false)$ to $O_i.(v_i,proposals_i,BC_i,\txD_i,oneTerm_i)$ (line~\ref{ln:vBasic}).
		
		\F
		\textbf{Argument (2)} \emph{Eventually $\mathrm{PROPOSAL}(v_i)$ messages are URB broadcast and $oneTerm_i$ holds.~~}
		By URB-termination, $\exist(O_i.\txD)$ does not hold eventually. Since $O_i.v \neq \bot$ (by the lemma assumption), the if-statement condition in line~\ref{ln:txDBotLandProp} holds and $p_i$ URB-broadcasts $\mathrm{PROPOSAL}(O_i.v)$. By applying again the same argument, the assignment in line~\ref{ln:vURBfix} makes sure that $oneTerm_i=\true$. 
		
		\F
		\textbf{Argument (3)} \emph{For any $p_x \in \sP: x \in \mathit{Correct}$, eventually $O_x.proposals[i] \neq \bot$ and $O_x.proposals[x] \neq \bot$ hold.~~}
		By URB-termination, every correct processor, $p_x$, eventually URB-delivers Argument (2)'s $\mathrm{PROPOSAL}(v_i)$ message. By the assumption that $v_i \neq \bot$ holds in any system state of $R$, the if-statement condition in line~\ref{ln:testSjVjNeqBot} holds (even if  $p_x$ has invoked $\mathsf{propose}_x(v_x)$ before this URB delivery). 
		
		In case there was no earlier invocation of $\mathsf{propose}_x(v_x)$, the assignment $O_x.v \gets v_i$ occurs due to line~\ref{ln:OgetsVj} (otherwise, a similar assignment occurs due to line~\ref{ln:proposalsGetsVj}). Moreover, due to the reasons that cause $p_i$ URB broadcasts in Argument (2), also $p_x$ URB broadcasts $\mathrm{PROPOSAL}(v' \neq \bot)$ messages. Upon the URB delivery of $p_x$ message to itself, the $O_x.proposals[x] \gets v' \neq \bot$ assignment occurs (line~\ref{ln:proposalsGetsVj}). (Note that this time, the if-statement condition in line~\ref{ln:proposalsGetsVj} must hold since $O_x\neq \bot$.) 
		
		\F
		\textbf{Argument (4)} \emph{The if-statement condition in lines~\ref{ln:proposalsIneqBotA} and~\ref{ln:proposalsIneqBotB} hold eventually.~~}
		Since $oneTerm_i$ holds eventually (Argument (2)), the if-statement condition in line~\ref{ln:proposalsIneqBotB} holds eventually. 
		Also, the fact that $O_x.proposals[x] \neq \bot$ (Argument (3)) and URB-termination imply that eventually, in $p_x$'s do-forever loop, the if-statement condition in 
		line~\ref{ln:proposalsIneqBotA} holds. In detail, since $R$ is an authentic execution, $k=\text{-}1 \land BC[k\text{+}1] = \bot$ holds in $R$'s second state, where $k=k_x()$. 

				\F
				
		Let  $S(z)=\{0,\ldots, z\}$. The proof of Argument (5) shows $\exists y\in S(n\text{-}1), \forall x \in \mathit{Correct}$, $p_x$ invokes $\mathsf{binPropose}_x()$ at most $y$ times and it observes that $O_x.BC[k]\neq \bot:k \in S(y\text{-}1)$. 
						
		\textbf{Argument (5)} \emph{The Termination property holds.~~}
		By line~\ref{ln:binProposeSeqProposalsB}, the if-statement condition in line~\ref{ln:proposalsIneqBotB} can hold at most once. The if-statement condition in line~\ref{ln:proposalsIneqBotA} cannot hold for more than $n$ times due to the $(k< n\text{-}1)$ clause. Thus, the termination property is implied. 

		\F
		%
		Let $r(j)=[x(0),\ldots,x(n\text{-}1)]:x(k)=BC_j[k].\done(k)$ and $S=\{[\bot,\ldots,\bot], [ \ldots,\false,\bot,$ $\ldots,\bot], [\ldots,\false,\true,\bot,\ldots,\bot], [\ldots,\false,$ $\true] \}$. For the case of using lines~\ref{ln:proposalsIneqBotA} to~\ref{ln:binProposeSeqProposalsA}, the proof of Argument (6) shows $\forall p_j \in \sP: r(j) \in S$, \ie sequential invocation of $\mathsf{binPropose}()$. 

		\F
		\textbf{Argument (6)} \emph{For the case of using lines~\ref{ln:proposalsIneqBotA} to~\ref{ln:binProposeSeqProposalsA}, $r(j)\in S$ holds.~~}
		%
%
		Due to lines~\ref{ln:KsMBC} and~\ref{ln:proposalsIneqBotA} as well as the agreement property of binary consensus objects and the fact that $R$ is authentic, we know that eventually, all non-failing nodes must observe the same results from their consensus objects. Specifically for the case of  lines~\ref{ln:proposalsIneqBotA} to~\ref{ln:binProposeSeqProposalsA}, it holds that $\forall p_j \in \sP: r(j)=r_s$. Also, at any time, $r(j)$ can only include a finite number (perhaps empty but with no more than $n\text{-}1$) of $\false$ values that are followed by at most one $\true$ value and the only $\bot$-values (if space is left), \ie $r(j)\in S$. 
		
		\F
		\textbf{Argument (7)} \emph{The Agreement property holds.~~}
		%
		
		Since no  $p_j \in \sP$ invokes $\mathsf{binPropose}_j(k_j\text{+}1,O_x.proposals[k_j\text{+}1] \neq \bot)$ (lines~\ref{ln:proposalsIneqBotA} and~\ref{ln:proposalsIneqBotB}), before it had assured the safe URB delivery of $O_x.\txD$'s transmission, we know that eventually, at least one element of $r(j)$ is $\true$. Thus, by the agreement property of binary consensus, every $p_x$ eventually calculates the same value of $k_j()$, such that $BC[k_j()].\done_x(k_j()+1)=\true$. This implies the agreement property since $\done_x()$ returns $O_x.proposals[k_j()\text{+}1]$ for any non-failing $p_x \in \sP$ (line~\ref{ln:returnedDecidedValue}).       
	\end{lemmaProof}
	
	\FF
	
	Lemma~\ref{thm:integrity} demonstrates the property of integrity.
	\begin{lemma}
		\label{thm:integrity}
		Suppose that $\exists v' \notin \{\bot,\blitza\}:\exists p_j \in \sP:\exists c'\in R: \done_j()=v'$ in $c'$. $\nexists c'' \in R: \done_j()=v''$ in $c''$, such that $v' \neq v''$.
	\end{lemma}


	\renewcommand{\lemcnt}{\ref{thm:integrity}}
	
	\begin{lemmaProof}
		The proof is by contradiction. 
		Suppose that $c'' \in R$ exists and, without the loss of generality, $c'$ appears before $c''$ in $R$. Since $R$ is authentic and $c' \in R$ exists, then there is a $p_k \in \sP:k \in S(n\text{-}1)$, such that for any $p_j \in \sP$, there is a system state $c'_j$ that appears in $R$ not after $c'$ in which for any $k' \in S(n\text{-}1)$ it holds that $BC[k'].\done_j(k') =\false$ for the case of $k'<k$ and $BC[k].\done_j(k') =\true$ for the case of $k'=k$. This is due the definition of $\mathsf{k}()$ (line~\ref{ln:curKs}).
		Note that in any system state that follows $c'_j$, the value of $k=\mathsf{k}_j()$ does not change due to the integrity of binary consensus objects. Therefore, $\done_j()$ must return the value of $O_j.proposal[k]$ in any system state that follows $c'_j$. Since line~\ref{ln:proposalsGetsVj} does not allow any change in the value of $O_j.proposal[k]$ between $c'$ and $c''$, it holds that $v'=v''$. Thus, the proof reached a contradiction and the lemma is true.
	\end{lemmaProof}
\end{theoremProof}


\Section{Application: Self-stabilizing Total-order Message Delivery using Multivalued Consensus}
\label{sec:bounded}
We exemplify the use of Algorithm~\ref{alg:consensus} by implementing the task total order uniform reliable broadcast (TO-URB), which we specified in Section~\ref{sec:TOURB}. We describe our implementation before bringing the correctness proof.


\begin{algorithm}[h!]
	\begin{\algSize}
		\noindent \textbf{notations:}\label{ln:xMod} $x~ \mathsf{opr}_\7~ y \equiv (x~ \mathsf{opr}~ y) \bmod \7:\mathsf{opr} \in \{\text{-},\text{+}\}$\; 
		
		\noindent \textbf{constants and variables:} 
		\remove{$M = 3$ number of multivalued consensus objects (Section~\ref{sec:spec});}
		$\delta \in \mathbb{Z}^+$ max number of messages after which is delivery is enforced;
		$CS[0..\6]$ $=[\bot,\ldots,\bot]$ array of multivalued consensus objects;
		$\xS=0$ highest obsolete sequence number\label{ln:varRxSTO} 
		
		\smallskip
		
		\textbf{macro} $\Sset()= \{CS[k].\mathit{seq}:CS[k]\neq\bot \}_{k \in \{0,\ldots,\6 \}}$\;
		
		\smallskip
		
		\textbf{macro} $\getSeq()$ \textbf{do} \Return{$\max(\{\xS\} \cup \Sset())$\label{ln:returnSeq}\;}
		
		\smallskip
		
		\textbf{macro} $\test(s)$ \textbf{do} \Return{$(s \in (\Sset() \cup \{\getSeq()+1\})$}\label{ln:textDo}\;
		
		\smallskip
		
		\textbf{macro} $\exceed()$ \textbf{do} \Return{$((\nonActive() \land 0<\ell)$ $\lor \delta \leq\ell)$} \textbf{where} $(x,y,\ell)=(\minReady(),$ $\maxReady(), \sum_{p_k \in \sP}(y[k]-x[k]))$\; 
		
		\smallskip
		
		\textbf{operation} $\mathsf{toBroadcast}(m)$ \textbf{do} \ems{$\mathbf{fifoURB}(\mathsf{toURB}(m))$}\label{ln:OptoBroadcast}; 
		
		\smallskip
		
		\textbf{do forever} \label{ln:doForEver}\Begin{
			
			\lIf{$(\exists k \in \{0,\ldots,\6 \} : CS[k]\neq \bot \land CS[k].\mathit{seq} \bmod \7 \neq k)\lor (\Sset()\neq \emptyset \land (\xS > \max \Sset() \lor$ $\max \Sset() - \min\Sset() > 1))$\label{ln:SneqEmptySetSeq}}{$CS \gets [\bot, \ldots ,\bot]$ \label{ln:CSgetsBots}}
			
			
			$sn \gets sn+1$\label{ln:anGetsAnPlusOne}\;
			
			\Repeat{$\mathrm{SYNCack}(sn, \bullet)$ \emph{received from \ems{all}} $p_j: j\in \mathit{trusted}$\label{ln:URBTOsendSYNCack}}{
				\lForEach{$p_j \in \sP$}{$\mathbf{send}~\mathrm{SYNC}(\mathit{sn})~ \mathbf{to}~ p_j$\label{ln:URBTOsendSYNC}}
			}
			
			\textbf{let} $(\ems{\mathit{allReady}},\mathit{maxSeq},\mathit{allSeq})=(\text{entrywise-min}\{x\}_{(\bullet,x) \in X}$, $\max \{x\}_{(\bull,x,\bullet) \in X}, \cup_{(\bull,x,y,\bull) \in X}\{x,y\})$\label{ln:letAgregate} \textbf{where} $X$ is the set of messages received in line~\ref{ln:URBTOsendSYNCack}\;
			
			\textbf{let} $(x,y,z)=(\xS,\getSeq(), \mathit{maxSeq})$\;
			
			\If{$\neg (x\text{+}1$$=$$y$$=$$z$$\lor$$x$$=$$y$$=z$$\lor$$x$$=$$y=$$z\text{-}1)$\label{ln:xyzGetSeq}}{$\xS \gets \max \{x,y,z\}$\label{ln:xyzGetSeqGet}}
			
			\lForEach{$k \in \{0,\ldots,\6\} \setminus ( \{x \bmod \7 : x < y\} \cup \{y$ $\bmod~ \7 \}  \cup \{z\text{+}_{\7}1 :|\mathit{allSeq}|=1\})$\label{ln:k06x7minAllSeq}}{$CS[k] \gets \bot$\label{ln:k06x7minAllSeqGet}}
			
			\If{$(|\mathit{allSeq}|=1\land \exceed())$\label{ln:allSeqeqoneexceedmaxSeqMinReady}}{
				$CS[\mathit{maxSeq}+_{\7}1].propose(\mathit{maxSeq}+1,\ems{\mathit{allReady}})$\label{ln:CSseqMpropose}
			}
			
			\If{$\xS+ 1 = \getSeq() \land x \neq \bot  \land x.\done() \neq \bot$ \textbf{\emph{where}} $x=CS[(\xS+_{\7}1)]$\label{ln:xNeqBotLand}}{{
					\If{$x.\done() \neq \blitza$}{\textbf{foreach }{$m \in \bulkRead(x.\done())$\label{ln:toDeliverMif}} \textbf{do} {$\mathsf{toDeliver}(m)$\label{ln:toDeliverM}}} 
				}
				$\xS \gets \xS+1$\label{ln:xSGetsxSPlusOne}
			}
		}
		
		\smallskip
		
		\textbf{upon} $\mathsf{SYNC}(\mathit{snJ})$ \textbf{arrival} \textbf{from} $p_j$ \textbf{do} 
		$\mathbf{send}~\mathrm{SYNCack}(\mathit{snJ}, getSeq(), \xS, \ems{\maxReady}())~ \mathbf{to}~ p_j$\label{ln:URBTOackSYNC}\;
		
		
		\caption{\label{alg:urbTO}Self-stabilizing TO-URB via consensus;  code for $p_i\in\sP$}	
		
	\end{\algSize}
	
\end{algorithm}

\Subsection{Refinement of the system settings}
\label{sec:modelRefined}
Our self-stabilizing total order message delivery implementation (Algorithm~\ref{alg:urbTO}) provides the $\mathsf{toBroadcast}(m)$ operation (line~\ref{ln:OptoBroadcast}). It uses a self-stabilizing URB with FIFO-order delivery, such as the one by Lundstr\"om, Raynal, and Schiller~\cite{selfStabURB}, to broadcast the protocol message,  $\mathsf{toURB}(msg)$. As before, the line numbers of Algorithm~\ref{alg:urbTO} continues the ones of Algorithm~\ref{alg:consensus}.

The proposed solution assumes that the FIFO-URB module has interface functions that facilitate the aggregation of protocol messages before their delivery. For example, we assume that the interface function $\nonActive_i()$ returns $\true$ whenever there are no active URB transmissions sent by $p_i \in \sP$. Also, given $p_i \in \sP$, the functions $\retrieve_i()$ and $\fifoReady_i()$ return each a vector, $r_i[0,..,n\text{-}1]$, such that for any $p_j \in \sP$, the entry $r_i[j]$ holds the lowest, and respectively, highest FIFO-delivery sequence number that is ready to be FIFO-delivered. These FIFO-delivery sequence numbers are the unique indices that the senders attach to the URB messages. 
Also, the function $\bulkRead_i(r_{\max})$ returns immediately after system state $c$ a determinately ordered sequence, $sqnc_i$, that includes all the messages between $r_{\min}$ and $r_{\max}$, such that $r_{\min}=\retrieve_i()$ in $c$, as well as $r_{\max}$, is a vector that is entry-wise greater or equal to $r_{\min}$ and entry-wise smaller or equal $r=\fifoReady_i()$ in $c$.

Algorithm~\ref{alg:urbTO} assumes access to a self-stabilizing perfect failure detector, such as the one by Beauquier and Kekkonen{-}Moneta~\cite{DBLP:journals/ijsysc/BeauquierK97}. The local set, $\mathit{trusted}_i$, includes the indices of the nodes that $p_i$'s failure detector trusts. We follow Assumption~\ref{thm:messageArrival} for the sake of a simple presentation.

\begin{assumption}
	\label{thm:messageArrival}
	Any sent message arrives or is lost within $\bigO(1)$ asynchronous cycles. Any URB message arrives within $\bigO(1)$ asynchronous cycles~\cite{selfStabURB}. Each active multivalued consensus object decides within $\bigO(1)$ asynchronous cycles~\cite{DBLP:conf/icdcn/LundstromRS21}.
\end{assumption}

\Subsection{Algorithm description}
\label{sec:algDesc}
The algorithm idea uses the fact that $sqnc_i$ is deterministically ordered. Namely, if all nodes $p_j \in \sP$ share the same sequence $r_1,r_2,\ldots$ when calling $\bulkRead_j(r_x):x\in \mathbb{Z}^+$, the studied task is reduced to invoking the event of $\mathsf{toDeliver}(m)$ for every $m \in \bulkRead(r_x)$. To that end, Algorithm~\ref{alg:urbTO} queries all nodes about the messages that are ready to be delivered (lines~\ref{ln:anGetsAnPlusOne} to~\ref{ln:letAgregate}), validates the consistency of the control variables (line~\ref{ln:SneqEmptySetSeq} and lines~\ref{ln:xyzGetSeqGet} to~\ref{ln:k06x7minAllSeqGet}), agree on the current value of $r_x$ (lines~\ref{ln:allSeqeqoneexceedmaxSeqMinReady} to~\ref{ln:CSseqMpropose}), and delivers the ready messages (lines~\ref{ln:xNeqBotLand} to~\ref{ln:xSGetsxSPlusOne}). We discuss in detail each part after describing the local constant, variables, and macros; $sn$ the query number.      

\Subsubsection{Constant, variables, and macros}
The constant $M$ defines the number of multivalued consensus objects that Algorithm~\ref{alg:urbTO} needs to use. The proof shows that, at any time during a legal execution, Algorithm~\ref{alg:urbTO} uses at most two active objects at a time and one more object that is always non-active. The array $CS[0..\6]$ holds all the multivalued consensus objects that Algorithm~\ref{alg:urbTO} uses. Algorithm~\ref{alg:urbTO} uses $CS[]$ cyclically.

Algorithm~\ref{alg:urbTO} aims at aggregating URB messages and delivering them only when all transmission activities have terminated. To that end, it uses the $\nonActive()$ function (Section~\ref{sec:modelRefined}). Since the number of such transmissions is unbounded, there is a need to stop aggregating after some predefined number of transmissions that we call $\delta$. The variable $\xS$ points to the highest obsolete sequence number; the one that was already delivered locally. The variable $sn$ stores the number of the next query. As mentioned in Section~\ref{sec:spec}, whenever Algorithm~\ref{alg:urbTO} runs out of query numbers, a global restart mechanism is invoked, such as the one in~\cite[Section~5]{DBLP:conf/netys/GeorgiouLS19}. Thus, it is possible to have bounded query numbers.

The macro $\Sset()$ returns the set of sequence numbers used by the active multivalued consensus objects. The macro $\getSeq()$ returns the locally maximum sequence number. The macro $\test(s)$ returns $\true$ whenever the sequence number $s$ is used by an active or is greater by one than $\getSeq()$. The macro $\exceed()$ facilitates the decision about whether to invoke a new consensus. It returns $\true$ if there are non-delivered messages but no on-going transmissions. It also returns $\true$ when the number of ready to be delivered messages exceeds $\delta$ (regardless of the presence of active URB transmissions).

\Subsubsection{Querying (lines~\ref{ln:anGetsAnPlusOne} to~\ref{ln:letAgregate})}
Algorithm~\ref{alg:urbTO} uses a simple synchronization query mechanism. Each query instance is associated with a unique sequence number that is stored in the variable $sn$ and incremented in line~\ref{ln:anGetsAnPlusOne}. Line~\ref{ln:URBTOsendSYNC} broadcasts the synchronization query repeatedly until a reply is received from every trusted node. The query response (line~\ref{ln:URBTOackSYNC}) includes the correspondent's (local) maximum sequence number stored by any multivalued consensus object (that the macro $\getSeq()$ retrieves), the maximum obsolete sequence number (that its respective multivalued consensus object is no longer needed), and the latest value returned from $\fifoReady_i()$. Using these responses, line~\ref{ln:URBTOsendSYNCack} aggregates the query results and store them in $\minReady$, $\mathit{maxSeq}$, and $\mathit{allSeq}$. The vector $\minReady$ includes the entry-wise minimum (per sender) FIFO-URB messages that are ready to be delivered at all nodes. Also, $\mathit{maxSeq}$ is the maximum known sequence number. And, the set $\mathit{allSeq}$ includes all the collected maximum sequence numbers and obsolete sequence numbers.            

\Subsubsection{Consistency assertion and stale information removal (line~\ref{ln:SneqEmptySetSeq} and lines~\ref{ln:xyzGetSeqGet} to~\ref{ln:k06x7minAllSeqGet})}
Line~\ref{ln:SneqEmptySetSeq} makes sure that $CS[k].\mathit{seq}$, when taken its reminder from the division by $M$, equals to $k$. It also tests that the local obsolete sequence number, $\xS$, is not greater than the largest sequence number. Besides, the gap between the maximum and the minimum sequence number cannot be greater than one. Line~\ref{ln:xyzGetSeqGet} verifies that $\xS$, $\getSeq()$, and $\mathit{maxSeq}$ follow a consistent pattern. Line~\ref{ln:k06x7minAllSeqGet} removes stale information by deactivating any obsolete multivalued consensus object. 


\Subsubsection{Repeated agreement (lines~\ref{ln:allSeqeqoneexceedmaxSeqMinReady} to~\ref{ln:CSseqMpropose})}
The if-statement condition in line~\ref{ln:allSeqeqoneexceedmaxSeqMinReady} tests whether all trusted nodes in the system share the same sequence number. This happens when all trusted nodes $p_j \in \sP$ have $\xS_j=\getSeq_j()$. Line~\ref{ln:allSeqeqoneexceedmaxSeqMinReady} also checks whether $\exceed()$ indicates that it is the time to deliver a batch of messages. If this is the case, then line~\ref{ln:CSseqMpropose} proposes to agree on the value of \ems{$\mathit{allReady}$.} 

\Subsubsection{Message delivery (lines~\ref{ln:xNeqBotLand} to~\ref{ln:xSGetsxSPlusOne})}
The delivery of the next message batch becomes possible the multivalued consensus object has terminated (line~\ref{ln:xNeqBotLand}). Before the actual delivery (line~\ref{ln:toDeliverMif}), there is a need to check that no error was reported (line~\ref{ln:toDeliverMif}) due to conditions that appear at \ems{line~\ref{ln:blitza}} of Algorithm~\ref{alg:consensus}. In any case, \ems{$\xS$} is incremented (line~\ref{ln:xSGetsxSPlusOne}) so that even if an error occurred, the object is ready for recycling.     

\Subsection{Correctness \ems{of Algorithm~\ref{alg:urbTO}}}
\label{sec:algDescCorr}
Theorem~\ref{thm:converTO} uses Definition~\ref{thm:consistentTO}.

\begin{definition}[Consistent states and legal executions]
	\label{thm:consistentTO}
	Let $c$ be a system state and $p_i \in \sP$ be any processor in the system. Suppose that in $c$, it holds that (i) $\forall k \in \{0,\ldots,\6 \} : CS_i[k]=\bot \lor CS_i[k].\mathit{seq} \bmod \7 = k$ and either  $S=\emptyset \land \xS_i \in \mathbb{Z}^+$ or $S \neq \emptyset \land (\xS_i \leq \max S \land \max S - \min S \leq \5)$, where $S=\{CS_i[k].\mathit{seq}:CS_i[k]\neq\bot \}_{k \in \{0,\ldots,\6 \}}$ and $\getSeq_i()$ returns $\mathit{seq}=\max(\{\xS_i\} \cup S)$. Moreover, (ii.a) $sn_i$'s value is greater equal to any $\mathit{snJ}$ field in the message $\mathsf{SYNC}(\mathit{snJ})$ in a communication channel from $p_i$ as well as $\mathrm{SYNCack}(\mathit{snJ},\bullet)$ message in a communication channel to $p_i$. And (ii.b) $\xS_i \leq \getSeq_i() \leq \xS_i+\5$. In this case, we say that $c$ is consistent concerning Algorithm~\ref{alg:urbTO}. 
	
	Suppose that $R$ is an execution of Algorithm~\ref{alg:urbTO}, such that every $c \in R$ is consistent. In addition, (iii.a) suppose that if $\mathsf{toBroadcast}()$ is not invoked during $R$ nor do any FIFO-broadcast becomes available for delivery, then $pred$ holds throughout $R$, where $pred \equiv \exists z \in \mathbb{Z}^+:\forall k \in \mathit{Correct}: \getSeq_k()=z \land \mathit{maxSeq}_k=z \land \xS_k=z\land \mathit{allSeq}_k=\{z\}$. Furthermore, (iii.b) suppose that if $\mathsf{toBroadcast}()$ is invoked during $R$ infinitely often, then $pred$ holds infinitely often. In this case, we say that $R$ is legal.
\end{definition}

\begin{theorem}
	\label{thm:converTO}
	Within $\bigO(1)$ asynchronous cycles, Algorithm~\ref{alg:urbTO}'s execution is legal.
	%
\end{theorem}
\renewcommand{\thmcnt}{\ref{thm:converTO}}

\begin{theoremProof}	
	Due to line~\ref{ln:CSgetsBots}, Definition~\ref{thm:consistentTO}'s Invariant (i) holds after $p_i$ first complete iteration of Algorithm~\ref{alg:urbTO}'s do-forever loop (lines~\ref{ln:doForEver} to~\ref{ln:xSGetsxSPlusOne}). Lemma~\ref{thm:sn} shows Invariant (ii.a). 
	%
	%
	Line~\ref{ln:xyzGetSeqGet} implies Invariant (ii.b). 	Lemma~\ref{thm:Q} shows invariant (iii). 
	
	\begin{lemma}
		\label{thm:sn}
		Invariant (ii.a) holds.
		\remove{Let $sn_i=x$ be the value of $p_i$'s $sn$ value in $R$'s starting system state. Within $\bigO(1)$ asynchronous cycles, $sn_i$'s value is greater equal to $x$ and any $\mathit{snJ}$ field in the message $\mathsf{SYNC}(\mathit{snJ})$ in a communication channel from $p_i$ as well as $\mathrm{SYNCack}(\mathit{snJ},\bullet)$ message in a communication channel to $p_i$.} 
	\end{lemma}
	\renewcommand{\lemcnt}{\ref{thm:sn}}
	
	\begin{lemmaProof}
		Only line~\ref{ln:anGetsAnPlusOne} modifies $sn_i$'s value\reduce{, \ie by increasing the value of $sn_i$}. The rest of the proof is implied by Assumption~\ref{thm:messageArrival}\reduce{, which says that all messages that appear in the communication channels in $R$'s starting system state are either delivered or lost within $\bigO(1)$ asynchronous cycles}.
	\end{lemmaProof}
	
	\FF
	We observe from the code of Algorithm~\ref{alg:consensus} that once invariants (i) and (ii) hold, they are not violated. Thus, the rest of the proof assumes that invariants (i) and (ii) hold in every system state of $R$. Lemma~\ref{thm:M} is needed for the proof of Lemma~\ref{thm:Q}.
	
	\begin{lemma}
		\label{thm:M}
		Every complete iteration of the do-forever loop (lines~\ref{ln:doForEver} to~\ref{ln:xSGetsxSPlusOne}) allows the collection of $M_{sn_i}=\{(sn_i,s_k,o_k,r_k)\}_{k \in \mathit{trusted}_i}$, such that $s_k=\mathit{seq}_k, o_k=\xS_k$, and $r_k=\fifoReady_k()$ in the system state $c_k$, where $c^{line~\ell}_i \in R:\ell \in \{\ref{ln:anGetsAnPlusOne},\ref{ln:URBTOsendSYNCack}\}$ is the system state when $p_i$ executed line $\ell$ with $sn_i$ and $c_k$ appears between $c^{line~\ref{ln:anGetsAnPlusOne}}_i$ and $c^{line~\ref{ln:URBTOsendSYNCack}}_i$ when $p_k$ executed line~\ref{ln:URBTOackSYNC} on the arrival of $\mathsf{SYNC}(\mathit{snJ=sn_i})$. Moreover, 
		%
		%
		\ems{$\minReady_i$ (line~\ref{ln:URBTOsendSYNCack}) is entry-wise smaller equal to every $\fifoReady_k()$ in $c^{line~\ref{ln:URBTOsendSYNCack}}_i$,  $\mathit{maxSeq}_i$ is greater equal than every $\mathit{seq}_k$ in $c^{line~\ref{ln:anGetsAnPlusOne}}_i$, and $\mathit{allSeq}_i$ includes the union $\cup_{k \in \mathit{trusted}_i} aS_k$, where $aS_k=\{\xS_k,\mathit{seq}_k\}$ in $c_k$.}
	\end{lemma}
\renewcommand{\lemcnt}{\ref{thm:M}}

	\begin{lemmaProof}
		Since invariants (ii.a) holds, the increment of $sn_i$ (line~\ref{ln:anGetsAnPlusOne}) creates a sequence number that is (associated with $p_i$ and) greater than all other associated sequence numbers in the system. With this unique sequence number, the repeat-until loop (lines~\ref{ln:URBTOsendSYNC} to~\ref{ln:URBTOsendSYNCack}) gets a fresh collection of $M_{sn_i}=\{(sn_i,s_k,o_k,r_k)\}_{k \in \mathit{trusted}_i}$. Note that this loop cannot block due to the end-condition (line~\ref{ln:URBTOsendSYNCack}), which considers only the trusted nodes\reduce{ in the system}. The rest of the proof is implied directly by lines~\ref{ln:anGetsAnPlusOne},~\ref{ln:URBTOsendSYNCack}, and~\ref{ln:URBTOackSYNC}. 
	\end{lemmaProof}
	
	
	\begin{lemma}
		\label{thm:Q}
		Within $\bigO(1)$ asynchronous cycles, $R=R'\circ R''$ reaches a suffix, $R''$, in which invariants (iii.a) and (iii.b) hold.
	\end{lemma}
	\renewcommand{\lemcnt}{\ref{thm:Q}}
	
	\begin{lemmaProof}	
		\textbf{Argument (1)} \emph{Invariant (iii.a) holds.~~}
		By the assumption that no FIFO-broadcast becomes ready during $R$, it holds that the if-statement condition in line~\ref{ln:allSeqeqoneexceedmaxSeqMinReady} does not hold during $R$. By Assumption~\ref{thm:messageArrival}, all active multivalued consensus objects have terminated with $\bigO(1)$ asynchronous cycles. Therefore, within $\bigO(1)$ asynchronous cycles, the if-statement condition in line~\ref{ln:toDeliverMif} cannot hold. Due to Lemma~\ref{thm:M}, $\mathit{maxSeq}_i$ is greater equal to $\getSeq_k():k \in \mathit{Correct}_i$. Due to the if-statement line~\ref{ln:xyzGetSeq} and line~\ref{ln:k06x7minAllSeq}, within $\bigO(1)$ asynchronous cycles, line~\ref{ln:k06x7minAllSeqGet} deactivates any multivalued consensus object, $O_{i:p_i \in \sP,x \in \{0,\ldots, \6\}}=CS_i[x]$ for which $O_{i,x}.\mathit{seq}<\mathit{maxSeq}_i-1$. By using Assumption~\ref{thm:messageArrival} again, any re-activated multivalued consensus object has to terminate with $\bigO(1)$ asynchronous cycles. Thus, the above implies that the state of all multivalued consensus objects, active or not, do not change and that $\mathit{maxSeq}_i=\getSeq_k():k \in \mathit{Correct}_i$. 
		
		We show that $\xS_i = \getSeq_i()$ holds within $\bigO(1)$ asynchronous cycles. By Invariant (ii.b), we know that either $\xS_i+ 1 = \getSeq_i()$ or $\xS_i = \getSeq_i()$.
		Suppose that $\xS_i+ 1 = \getSeq_i()$ holds. Due to the definition of $\getSeq()$ as well as lines~\ref{ln:CSgetsBots} and~\ref{ln:k06x7minAllSeqGet}, $x_i \neq \bot$, where $x_i=CS_i[(\xS_i+_{\7}1)]$ (line~\ref{ln:xNeqBotLand}). By Assumption~\ref{thm:messageArrival}, within $\bigO(1)$ asynchronous cycles, the multivalued consensus object $x_i$ terminates. Thus, the if-statement condition in line~\ref{ln:xNeqBotLand} holds and line~\ref{ln:xSGetsxSPlusOne} increments $\xS_i$ once. Therefore, $\xS_i = \getSeq_i()$ within $\bigO(1)$ asynchronous cycles.
		
		Since $\mathit{maxSeq}_i=\getSeq_k()=\xS_k :k \in \mathit{Correct}_i$, then $\mathit{allSeq}_k=\{z\}$, where $z=\mathit{maxSeq}_i=\getSeq_k()=\xS_k :k \in \mathit{Correct}_i$. Thus, $pred$ holds.
		
		
		\F
		\textbf{Argument (2)} \emph{Invariant (iii.b) holds.~~}
		We note that $\exceed_i()$ holds infinitely often by the assumption that $\mathsf{toBroadcast}()$ is invoked during $R$ infinitely often and the URB-termination property. 
		We show that the if-statement condition in line~\ref{ln:allSeqeqoneexceedmaxSeqMinReady} holds within $\bigO(1)$ asynchronous cycles once $\exceed_i()$ holds. Suppose, towards a contradiction, that $|\mathit{allSeq}|=1$ does not hold for a period longer than $\bigO(1)$ asynchronous cycles. Then, the then-statement in line~\ref{ln:CSseqMpropose} is not executed for a period longer than $\bigO(1)$ asynchronous cycles. By the proof of Argument (1), $pred$ holds within $\bigO(1)$ asynchronous cycles. Thus, the if-statement condition in line~\ref{ln:allSeqeqoneexceedmaxSeqMinReady} holds within $\bigO(1)$ asynchronous cycles.	In other words, Invariant (iii.b) holds.
	\end{lemmaProof}
\end{theoremProof}

\begin{algorithm}[t!]
	
	\begin{\algSize}
		
		\smallskip
		\pushline\dosemic\nonl 
		
		
		\emph{Same code as in lines~\ref{ln:xMod} to~\ref{ln:OptoBroadcast}.}\\
		
		\popline
		
		\setcounter{AlgoLine}{51}
		
		\textbf{do forever} \Begin{
			
			\pushline\dosemic\nonl 
			
			\emph{Same code as in lines~\ref{ln:SneqEmptySetSeq} to~\ref{ln:k06x7minAllSeqGet}.}\\
			
			\popline

			\setcounter{AlgoLine}{61}
			
			\If{$(|\mathit{allSeq}|=1\land \exceed())$\label{ln:vallSeqeqoneexceedmaxSeqMinReady}}{
				$CS[\mathit{maxSeq}+_{\7}1].propose(\mathit{maxSeq}+1,(state=\mathsf{getState}(),msg=\mathit{maxReady())})$\label{ln:vCSseqMpropose}
			}
			
			\If{$\xS+ 1 = \getSeq() \land x \neq \bot  \land x.\done() \neq \bot$ \textbf{\emph{where}} $x=CS[(\xS+_{\7}1)]$\label{ln:vxNeqBotLand}}{{
					\If{$x.\done() \neq \blitza$}{$\mathsf{setState}(x.\done().state)$; ~~~~~~~~~~~ \textbf{foreach }{$m \in \bulkRead(x.\done().msg)$\label{ln:vtoDeliverMif}} \textbf{do} {$\mathsf{toDeliver}(m)$\label{ln:vtoDeliverM}} ~~~~~~~~ ~~~~~ ~~~~ ~~~~ ~~~~ \emph{Same code as in lines~\ref{ln:xSGetsxSPlusOne} to~\ref{ln:URBTOackSYNC}.}} 
				}
				

			}
		}
		
		%
		%
		
		\caption{\label{alg:urbTOV}Self-stabilizing emulation of a replicated state-machine;  code for $p_i\in\sP$}	
		
	\end{\algSize}
	
\end{algorithm}

\Section{Discussion}
%
%
\label{sec:disc}
We showed how a non-self-stabilizing algorithm for multivalued consensus by Most{\'{e}}faoui, Raynal, and Tronel~\cite{DBLP:journals/ipl/MostefaouiRT00} can become one that recovers from transient-faults. Interestingly, our solution is both wait-free and incurs a bounded number of binary consensus invocations whereas earlier work either uses an unbounded number of binary consensus objects~\cite{DBLP:journals/ipl/MostefaouiRT00} or is blocking~\cite{DBLP:journals/ipl/ZhangC09a}. \ems{Therefore, we present a more attractive transformation technique than the studied algorithm (regardless of the presence or absence of transient-faults).} 

\ems{As an application, we showed a self-stabilizing total-order message delivery. As an enhancement to this application, Algorithm~\ref{alg:urbTOV} explains how to construct a self-stabilizing emulator for state-machine replication. Line~\ref{ln:vCSseqMpropose} of Algorithm~\ref{alg:urbTOV} proposes to agree on both on the automaton state, which is retrieved by $\mathsf{getState}()$, and the bulk of FIFO-URB messages, as in line~\ref{ln:CSseqMpropose} of Algorithm~\ref{alg:urbTO}. Line~\ref{ln:vtoDeliverM} of Algorithm~\ref{alg:urbTOV} uses $\mathsf{setState}()$ for updating the automaton state using the agreed state.} 



\end{document}